\documentclass[aps,prl,twocolumn,showpacs,superscriptaddress,groupedaddress]{revtex4} 
\usepackage{graphicx}  
\usepackage{dcolumn}   
\usepackage{bm}        
\usepackage{amssymb}   
\usepackage{amsmath}
\usepackage{epsfig}
\usepackage{lipsum}
\usepackage{lineno}
\usepackage{hhline}
\begin{document}
\widetext

\newcommand*{\ANL}{Argonne National Laboratory, Argonne, Illinois 60439}
\newcommand*{\ANLindex}{1}
\affiliation{\ANL}
\newcommand*{\ASU}{Arizona State University, Tempe, Arizona 85287-1504}
\newcommand*{\ASUindex}{2}
\affiliation{\ASU}
\newcommand*{\CSUDH}{California State University, Dominguez Hills, Carson, CA 90747}
\newcommand*{\CSUDHindex}{3}
\affiliation{\CSUDH}
\newcommand*{\CANISIUS}{Canisius College, Buffalo, NY}
\newcommand*{\CANISIUSindex}{4}
\affiliation{\CANISIUS}
\newcommand*{\CMU}{Carnegie Mellon University, Pittsburgh, Pennsylvania 15213}
\newcommand*{\CMUindex}{5}
\affiliation{\CMU}
\newcommand*{\CUA}{Catholic University of America, Washington, D.C. 20064}
\newcommand*{\CUAindex}{6}
\affiliation{\CUA}
\newcommand*{\SACLAY}{IRFU, CEA, Universit'e Paris-Saclay, F-91191 Gif-sur-Yvette, France}
\newcommand*{\SACLAYindex}{7}
\affiliation{\SACLAY}
\newcommand*{\UCONN}{University of Connecticut, Storrs, Connecticut 06269}
\newcommand*{\UCONNindex}{8}
\affiliation{\UCONN}
\newcommand*{\DUQU}{Duquesne University, Pittsburgh, Pennsylvania. 15282  USA}
\newcommand*{\DUQUindex}{9}
\affiliation{\DUQU}
\newcommand*{\FU}{Fairfield University, Fairfield CT 06824}
\newcommand*{\FUindex}{10}
\affiliation{\FU}
\newcommand*{\FIU}{Florida International University, Miami, Florida 33199}
\newcommand*{\FIUindex}{11}
\affiliation{\FIU}
\newcommand*{\FSU}{Florida State University, Tallahassee, Florida 32306}
\newcommand*{\FSUindex}{12}
\affiliation{\FSU}
\newcommand*{\Genova}{Universit$\grave{a}$ di Genova, 16146 Genova, Italy}
\newcommand*{\Genovaindex}{13}
\affiliation{\Genova}
\newcommand*{\GWUI}{The George Washington University, Washington, DC 20052}
\newcommand*{\GWUIindex}{14}
\affiliation{\GWUI}
\newcommand*{\INFNFE}{INFN, Sezione di Ferrara, 44100 Ferrara, Italy}
\newcommand*{\INFNFEindex}{15}
\affiliation{\INFNFE}
\newcommand*{\INFNFR}{INFN, Laboratori Nazionali di Frascati, 00044 Frascati, Italy}
\newcommand*{\INFNFRindex}{16}
\affiliation{\INFNFR}
\newcommand*{\INFNGE}{INFN, Sezione di Genova, 16146 Genova, Italy}
\newcommand*{\INFNGEindex}{17}
\affiliation{\INFNGE}
\newcommand*{\INFNRO}{INFN, Sezione di Roma Tor Vergata, 00133 Rome, Italy}
\newcommand*{\INFNROindex}{18}
\affiliation{\INFNRO}
\newcommand*{\INFNTUR}{INFN, Sezione di Torino, 10125 Torino, Italy}
\newcommand*{\INFNTURindex}{19}
\affiliation{\INFNTUR}
\newcommand*{\INFNPAV}{INFN, Sezione di Pavia, 27100 Pavia, Italy}
\newcommand*{\INFNPAVindex}{20}
\affiliation{\INFNPAV}
\newcommand*{\ORSAY}{Institut de Physique Nucl'eaire, IN2P3-CNRS, Universit'e Paris-Sud, Universit'e Paris-Saclay, F-91406 Orsay, France}
\newcommand*{\ORSAYindex}{21}
\affiliation{\ORSAY}
\newcommand*{\JMU}{James Madison University, Harrisonburg, Virginia 22807}
\newcommand*{\JMUindex}{22}
\affiliation{\JMU}
\newcommand*{\KNU}{Kyungpook National University, Daegu 41566, Republic of Korea}
\newcommand*{\KNUindex}{23}
\affiliation{\KNU}
\newcommand*{\LAMAR}{Lamar University, 4400 MLK Blvd, PO Box 10009, Beaumont, Texas 77710}
\newcommand*{\LAMARindex}{24}
\affiliation{\LAMAR}
\newcommand*{\MISS}{Mississippi State University, Mississippi State, MS 39762-5167}
\newcommand*{\MISSindex}{25}
\affiliation{\MISS}
\newcommand*{\ITEP}{National Research Centre Kurchatov Institute - ITEP, Moscow, 117259, Russia}
\newcommand*{\ITEPindex}{26}
\affiliation{\ITEP}
\newcommand*{\UNH}{University of New Hampshire, Durham, New Hampshire 03824-3568}
\newcommand*{\UNHindex}{27}
\affiliation{\UNH}
\newcommand*{\NSU}{Norfolk State University, Norfolk, Virginia 23504}
\newcommand*{\NSUindex}{28}
\affiliation{\NSU}
\newcommand*{\OHIOU}{Ohio University, Athens, Ohio  45701}
\newcommand*{\OHIOUindex}{29}
\affiliation{\OHIOU}
\newcommand*{\ODU}{Old Dominion University, Norfolk, Virginia 23529}
\newcommand*{\ODUindex}{30}
\affiliation{\ODU}
\newcommand*{\RPI}{Rensselaer Polytechnic Institute, Troy, New York 12180-3590}
\newcommand*{\RPIindex}{31}
\affiliation{\RPI}
\newcommand*{\ROMAII}{Universita' di Roma Tor Vergata, 00133 Rome Italy}
\newcommand*{\ROMAIIindex}{32}
\affiliation{\ROMAII}
\newcommand*{\MSU}{Skobeltsyn Institute of Nuclear Physics, Lomonosov Moscow State University, 119234 Moscow, Russia}
\newcommand*{\MSUindex}{33}
\affiliation{\MSU}
\newcommand*{\SCAROLINA}{University of South Carolina, Columbia, South Carolina 29208}
\newcommand*{\SCAROLINAindex}{34}
\affiliation{\SCAROLINA}
\newcommand*{\TEMPLE}{Temple University,  Philadelphia, PA 19122 }
\newcommand*{\TEMPLEindex}{35}
\affiliation{\TEMPLE}
\newcommand*{\JLAB}{Thomas Jefferson National Accelerator Facility, Newport News, Virginia 23606}
\newcommand*{\JLABindex}{36}
\affiliation{\JLAB}
\newcommand*{\UTFSM}{Universidad T\'{e}cnica Federico Santa Mar\'{i}a, Casilla 110-V Valpara\'{i}so, Chile}
\newcommand*{\UTFSMindex}{37}
\affiliation{\UTFSM}
\newcommand*{\BRESCIA}{Universit`a degli Studi di Brescia, 25123 Brescia, Italy}
\newcommand*{\BRESCIAindex}{38}
\affiliation{\BRESCIA}
\newcommand*{\GLASGOW}{University of Glasgow, Glasgow G12 8QQ, United Kingdom}
\newcommand*{\GLASGOWindex}{39}
\affiliation{\GLASGOW}
\newcommand*{\YORK}{University of York, York YO10 5DD, United Kingdom}
\newcommand*{\YORKindex}{40}
\affiliation{\YORK}
\newcommand*{\VIRGINIA}{University of Virginia, Charlottesville, Virginia 22901}
\newcommand*{\VIRGINIAindex}{41}
\affiliation{\VIRGINIA}
\newcommand*{\VTECH}{Virginia Polytechnic Institute and State University, Blacksburg, Virginia 24061, USA}
\newcommand*{\VTECHindex}{42}
\affiliation{\VTECH}
\newcommand*{\WM}{College of William and Mary, Williamsburg, Virginia 23187-8795}
\newcommand*{\WMindex}{43}
\affiliation{\WM}
\newcommand*{\YEREVAN}{Yerevan Physics Institute, 375036 Yerevan, Armenia}
\newcommand*{\YEREVANindex}{44}
\affiliation{\YEREVAN}

\newcommand*{\NOWISU}{Idaho State University, Pocatello, Idaho 83209}
\newcommand*{\NOWJLAB}{Thomas Jefferson National Accelerator Facility, Newport News, Virginia 23606}
\newcommand*{\NOWINFNGE}{INFN, Sezione di Genova, 16146 Genova, Italy}

\title{Exclusive $\bm{\pi^{0}p}$ electroproduction off protons in the
  resonance region at photon virtualities 0.4~GeV$\bm{^{2}}$ $\bm{\leq~ Q^{2} \leq~1}$~GeV$\bm{^{2}}$}
\author{N. Markov}
\email{markov@jlab.org}
\affiliation{\UCONN}
\affiliation{\JLAB}
\author{K. Joo}
\affiliation{\UCONN}
\author{V.D. Burkert}
\affiliation{\JLAB}
\author{V.I. Mokeev}
\affiliation{\JLAB}
\author{ L. C. Smith}
\affiliation{\VIRGINIA}
\author{M. Ungaro}
\affiliation{\JLAB}

\author {S. Adhikari} 
\affiliation{\FIU}
\author {M.J.~Amaryan} 
\affiliation{\ODU}
\author {G. Angelini} 
\affiliation{\GWUI}
\author {H.~Atac} 
\affiliation{\TEMPLE}
% ...

\collaboration{The CLAS Collaboration}
\noaffiliation

\newcommand*{\dsig}{d\sigma/d\Omega_{\pi^0}}

\begin {abstract}

The exclusive electroproduction process $ep \rightarrow e'p'\pi^{0}$ was measured in the range of  photon virtualities $Q^{2} = 0.4 - 1.0$~GeV$^{2}$ and the invariant mass range of the $p\pi^{0}$ system of $W = 1.1 - 1.8$~GeV. These kinematics are covered in exclusive $\pi^{0}$ electroproduction off the proton with nearly complete angular coverage in the $p\pi^{0}$ center-of-mass system and with high statistical accuracy. Nearly 36000 cross section points were measured, and the structure functions $\sigma_T+\epsilon\sigma_L$,  $\sigma_{LT}$, and $\sigma_{TT}$,  were extracted via fitting the $\phi_{\pi^{0}}$ dependence of the cross section. A Legendre polynomial expansion analysis demonstrates the sensitivity of our data to high-lying $N^*$ and $\Delta^{*}$ resonances with $M~>~1.6$ GeV. As part of a broad effort to determine the electrocouplings of the $N^{*}$ and $\Delta^{*}$ resonances using both single- and double-pion electroproduction, this dataset is crucial for the reliable extraction of the high-lying resonance electrocouplings from the combined isospin analysis of the $N \pi$ and $\pi^{+}\pi^{-} p$ channels.
\end{abstract}
\maketitle
\section{Introduction}

The excitation of nucleon resonances via the electromagnetic interaction is an important source of information on the structure of excited nucleon states and dynamics of the non-perturbative strong interaction underlying the resonance formation \cite{aznaryanBurkert,BurkertRoberts}. The nucleon resonance electroexcitation amplitudes ($\gamma_vpN^*$ electrocouplings) are the primary source of information on many facets of non-perturbative strong interactions in the generation of the excited proton states with different structural features. Detailed studies of resonance electroexcitation in exclusive meson electroproduction off nucleons became feasible only after dedicated experiments were carried out with the CLAS detector~\cite{Burkert17} in Hall B at Jefferson Lab. CLAS produced the dominant part of the world exclusive meson electroproduction data in the nucleon resonance region and in the range of photon virtuality  $Q^2$ up to 5.0~GeV$^2$.  The data are available in the CLAS Physics Database \cite{clasBD}. Analyses of these data provided information on electrocouplings of most excited nucleon states in the mass range up to 1.8 GeV and at photon virtualities $Q^2$ $<$~5.0~GeV$^2$ \cite{Mokeev17}. The results on $\gamma_vpN^*$ electrocouplings are available at the web sites \cite{webIsupov,isupov-web}. 

The most detailed information on the $Q^2$-evolution of the $\gamma_vpN^*$ electrocouplings is available for the excited nucleon states in the mass range up to 1.6~GeV. These states couple preferentially to the $N\pi$ final states. Exclusive $N\pi$ electroproduction is the major source of information about their electrocouplings \cite{Laveissiere:2004, Kelly:2005,Kelly:2007,Smith:2006,JooSmith:2002,Ungaro:2006, Frolov:1999}.  The $\gamma_vpN^*$ electrocouplings of the resonances with masses $<$ 1.6~GeV were determined from independent studies of N$\pi$ \cite{JANR,aznauryan}, N$\eta$ \cite{Den07} and $\pi^+\pi^-p$ \cite{Mo09,mokeev2Pion,mokeev21Pion} electroproduction off protons. Consistent results on these resonance electrocouplings from independent analyses of different exclusive meson electroproduction channels support the available data on these fundamental quantities. The $\gamma_vpN^*$ electrocouplings of several nucleon resonances determined from the CLAS measurements are included in the recent PDG edition \cite{pdg}.

These data have a profound impact on our understanding of active degrees of freedom in the N$^*$ structure and the strong QCD dynamics underlying the generation of excited nucleon states. Analysis of the results on $\gamma_vpN^*$ electrocouplings within modern theoretical approaches with traceable connection to the QCD Lagrangians, such as Dyson-Schwinger Equation (DSE)~\cite{BurkertRoberts,Seg14,Seg15} and the combination of Light Cone Sum Rule (LCSR) and Lattice QCD~\cite{Br15, Br14} as well Light Front Relativistic Quarks models~\cite{aznuryan12, aznuryan15, aznuryan151, aznuryan17,obukh12,lub17,santopinto}  revealed the $N^*$ structure as a complex interplay between inner core of three dresses quarks and external meson-baryon cloud. The DSE approach~\cite{Seg14, Seg15} provided good descriptions of $\Delta(1232)3/2^+$ and N(1440)$1/2^+$ electrocouplings at $Q^2$~$>$~2.0~GeV$^2$ starting from the QCD Lagrangian and and shed light on the strong QCD dynamics, underlying the dominant part of hadron mass generation. Possibility to explore the hadron mass generation was demonstrated in conceptually different analyses of experimental results on electrocouplings of many resonances in the mass range up to 1.7~GeV carried out within the novel relativistic quark models~\cite{aznuryan12, aznuryan15, aznuryan151, aznuryan17}. 

The CLAS Collaboration keeps gradually extending the kinematic coverage of the experimental data on $\pi^+n$, $\pi^0p$, and $\pi^+\pi^-p$ photo- and electroproduction off protons over $W$ and $Q^2$ \cite{Gol18,parkAznauryan,Is17}. The  $\pi^+$n data in the third resonance region \cite{parkAznauryan} allowed us to determine electrocouplings of $N(1675)5/2^-$, $N(1680)5/2^+$, $N(1710)1/2^+$ resonances at 2.0~GeV$^2$~$<$~$Q^2$~$<$~5.0~GeV$^2$. Data on $\pi^0p$ electroproduction off proton available so far \cite{JooSmith:2002,lcs07,Bis08} were used mostly for studies of the $\Delta(1232)3/2^+$ electroexcitation amplitudes \cite{aznauryan} because of the limited statistical and systematical accuracy of these data in the mass range above the first resonance region. The combined studies of $\pi^+$n and $\pi^0$p electroproduction off protons are of particular importance for the extraction of both $\Delta^*$ and $N^*$ electrocouplings. The $\pi^0$p electroproduction channels offer preferential opportunities for the exploration of the $\Delta^*$ resonances because of the isospin Clebsch-Gordan coefficient values which enter in their hadronic decay amplitudes to the $\pi^+n$ and $\pi^0p$ final states.

The new precise data set of $\pi^0 p$ differential cross sections off protons presented in this paper cover the range of the W from 1.1~GeV to 1.8~GeV  at photon virtualities from 0.4~GeV$^{2}$ to 1~GeV$^{2}$. These new $\pi^0p$ data are essential in order to obtain electrocouplings of many resonances in the mass range from 1.5~GeV to 1.75~GeV contributing to N$\pi$ electroproduction off protons. In this paper, we demonstrate this in exploratory studies of the $\pi^0p$ data sensitivity to the variation of the resonance electrocouplings available from the previous results \cite{Mokeev17,webIsupov,isupov-web}. Recently, new data on exclusive $\pi^+\pi^-p$ electroproduction were published \cite{Fe18}. These data were obtained from the same experimental run as $\pi^0$p electroproduction off proton data presented in this paper and with the same coverage over $W$ and $Q^{2}$.

This paper is organized as follows: the general reaction formalism is outlined and followed by a brief description of the experimental setup and data taking. Charged particle identification is defined along with the selection of the fiducial regions for both, electron ($e$) and proton ($p$).  Event selection is completed by the identification of the $\pi^{0}$ using the missing mass technique and reaction kinematics. Corrections for acceptance, radiative effects, empty target, and bin centering are developed and applied to the raw event yields. The absolute normalization is checked against benchmark reactions and the major sources of systematic errors are identified.  Cross sections and structure functions are compared with model predictions in different $W$ regions and resonance contribution into the cross section is estimated. Legendre polynomials are extracted and show the sensitivity of the obtained data to selected nucleon resonant states.
\section{Formalism}
\begin{figure}[htp]
\begin{center}
\includegraphics[width=6.5cm, angle=270]{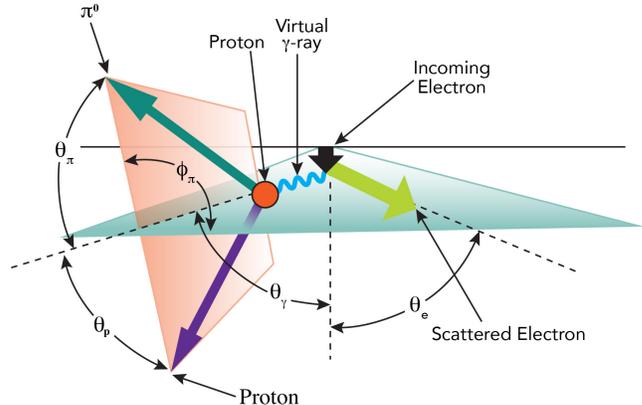}
\caption{Schematics of single $\pi^{0}$ electroproduction.}
\label{reactionScheme}
\end{center}
\end{figure}

The schematics of $\pi^{0}$ electroproduction off the proton are presented in Fig.~\ref{reactionScheme}, where the incoming electron $e$ emits a virtual photon $\gamma^{*}$, which is absorbed by the target proton $p$. The incoming and outgoing electron form the scattering plane, while the recoiling proton and $\pi^{0}$ form the reaction plane.  The direction of the outgoing pion is determined by the angle $\phi_{\pi^{0}}$ between these planes and the angle $\theta_{\pi^{0}}$ between the direction of the pion and the virtual photon. The virtual photon is described by the value of the photon virtuality $Q^{2}$, energy transfer $\nu$, and polarization $\epsilon$:
\begin{eqnarray}
\nu &=& E_{i} - E_{f},\\
Q^{2} &=& 4E_{i}E_{f}\mathrm{sin}^{2}\frac{\theta_{e}}{2}, \mathrm{and}\\
\epsilon &=& \frac{1}{1 + 2(1 + \frac{\nu^{2}}{Q^{2}}\mathrm{tan}^{2}\frac{\theta_{e}}{2})},
\end{eqnarray}
where $E_{i}$ and $E_{f}$ are the initial and final energy of the electron and  $\theta_{e}$ is the polar angle of the scattered electron with respect to the incoming electron.
The $(e,e')X$ missing mass $M_X$ (denoted as W throughout the text) is
\begin{equation}
W = \sqrt{M_{p}^{2} + 2M_{p}\nu - Q^{2}},
\end{equation}
where $M_{p}$ is the mass of the proton.
In the one-photon-exchange approximation, the four-fold differential cross section of $\pi^{0}$ electroproduction relates to $\frac{d\sigma}{d\Omega_{\pi^{0}}}$, as
\begin{equation}
\frac{d^{4}\sigma}{dWdQ^{2}d\Omega_{\pi^{0}}} =
J\Gamma_\nu\frac{d\sigma}{d\Omega_{\pi^{0}}},
\label{csFluxEquation}
\end{equation}
where the Jacobian 
\begin{equation}
J = \frac{\partial(Q^2,W)}{\partial(E_{f},\cos\,\theta_e,\phi_e)} = \frac{2ME_{i}E_{f}}{W}
\end{equation}
 relates the differential
volume element $dQ^2 dW$ of the binned data to the measured electron kinematics 
$dE_{f}\,d\cos\theta_e\,d\phi_e$ and $\Gamma_\nu$ is the virtual photon flux,
\begin{equation}
\Gamma_\nu = \frac{\alpha}{2\pi^{2}}\frac{E_{f}}{E_{i}}\frac{k_{\gamma}}{Q^{2}}\frac{1}{1-\epsilon},
\end{equation}
where $\alpha$ is the fine structure constant and $k_{\gamma}  = \frac{W^{2} -
  m_{p}^{2}}{2m_{p}}$ is the photon equivalent energy. 
  Assuming single photon exchange for the description of exclusive $\pi^0p$ electroproduction, the expression for $\dsig$ can be written as
\begin{widetext}
\begin{equation}
\frac{d\sigma}{d\Omega_{\pi^0}} = \frac{p_{\pi^0}}{k^*_{\gamma}}((\sigma_T+\epsilon \sigma_L) + \sigma_{LT}\sqrt{2\epsilon(\epsilon+1)}\mathrm{sin}\theta_{\pi^{0}}\mathrm{cos}\phi_{\pi^{0}} + \epsilon \sigma_{TT}\mathrm{sin}^{2}\theta_{\pi^{0}} \mathrm{cos2}\phi_{\pi^{0}}), 
\label{crossSectionFormula}
\end{equation}
\end{widetext}
where $p_{\pi^0}$, $\theta_{\pi^{0}}$, and $\phi_{\pi^{0}}$ are the absolute values of the three-momentum, polar and azimuthal angles of the ${\pi^0}$ in the CM frame, and $k^*_{\gamma}=k_{\gamma}m_p/W$. 

From Eq. (\ref{crossSectionFormula}), the combination $\sigma_T+\epsilon \sigma_L$ is determined by the modulus squared of the single pion electroproduction amplitudes. The two other terms represent the interference structure functions, namely, $\sigma_{TT}$ describes the interference between amplitudes with transversely polarized virtual photons of +1 and -1 helicities, while $\sigma_{LT}$ is determined by the interference between amplitudes with a longitudinal virtual photon of helicity 0 and the difference of the two transverse photon amplitudes of helicities +1 and -1~\cite{Am72}.
\section{Experimental setup}
This experiment used the CEBAF Large Acceptance Spectrometer (CLAS)~\cite{CLASDetector} in Hall B at Jefferson Lab. The detector is divided into six independent identical spectrometers (referred to as sectors), and has a nearly $4\pi$ angular coverage in the center-of-mass system, which makes it ideally suited for experiments that require detection of several particles in the final state.
A toroidal magnetic field created by six superconducting coils around the beam line bends the trajectories of the charged particles to measure their momentum using Drift Chambers (DC) ~\cite{dcMestayer}, while scintillator counters (SC) ~\cite{tofSmith} are used to measure their time of flight. Gas threshold Cherenkov Counters (CC)~\cite{ccAdams} are used for the separation of electrons from negative pions. Electromagnetic Calorimeters (EC) uses a lead-scintillator sandwich design~\cite{ecAmarian} samples the electromagnetic showers to identify electrons and also to provide neutral particle detection.

A 2 cm long cryogenic liquid hydrogen ($LH_{2}$) target cell is located near the center of the setup, surrounded by a small mini-torus magnet used to deflect low-energy M$\o$ller electrons out of the CLAS acceptance. A Faraday cup installed at the end of the beam line measured the full beam charge passing through the target.
\section{Data taking}
The data reported in this analysis were taken during the e1e run period in Hall B in the period of November 2002 - January 2003. A longitudinally polarized electron beam with energy of 2.036~GeV was incident on the target. The torus current was set at 2250A, and the mini-torus current was 5995 A. The nominal beam current during the run was set at 10 nA. The total charge accumulated for the runs used in the analysis was 6 mC. Several empty target runs were performed to estimate the contribution from the target entry and exit windows.

The event readout was triggered by the coincidence of  signals from the Electromagnetic Calorimeter and Cherenkov counters in the same sector. The total number of accumulated triggers was $\sim$ $10^{9}$. 
\section{Particle identification}\label{particleID}
\subsection{Electron identification}
An electron candidate requires a negatively charged track in the DC matched to a hit both in the CC and EC detectors. The EC is used to trigger on electromagnetic showers generated by electrons, and to reject minimum-ionizing particles, such as pions, which deposit a constant amount of energy per unit path travelled through the scintillator material. For particles that hit the calorimeter near its edge, the shower produced may not have been fully contained within the calorimeter. Therefore these border regions of the calorimeter are eliminated using geometrical fiducial cuts applied on the cluster hit coordinates in the calorimeter.

\begin{figure}[htb]
  \begin{center}
    \epsfig{file =
   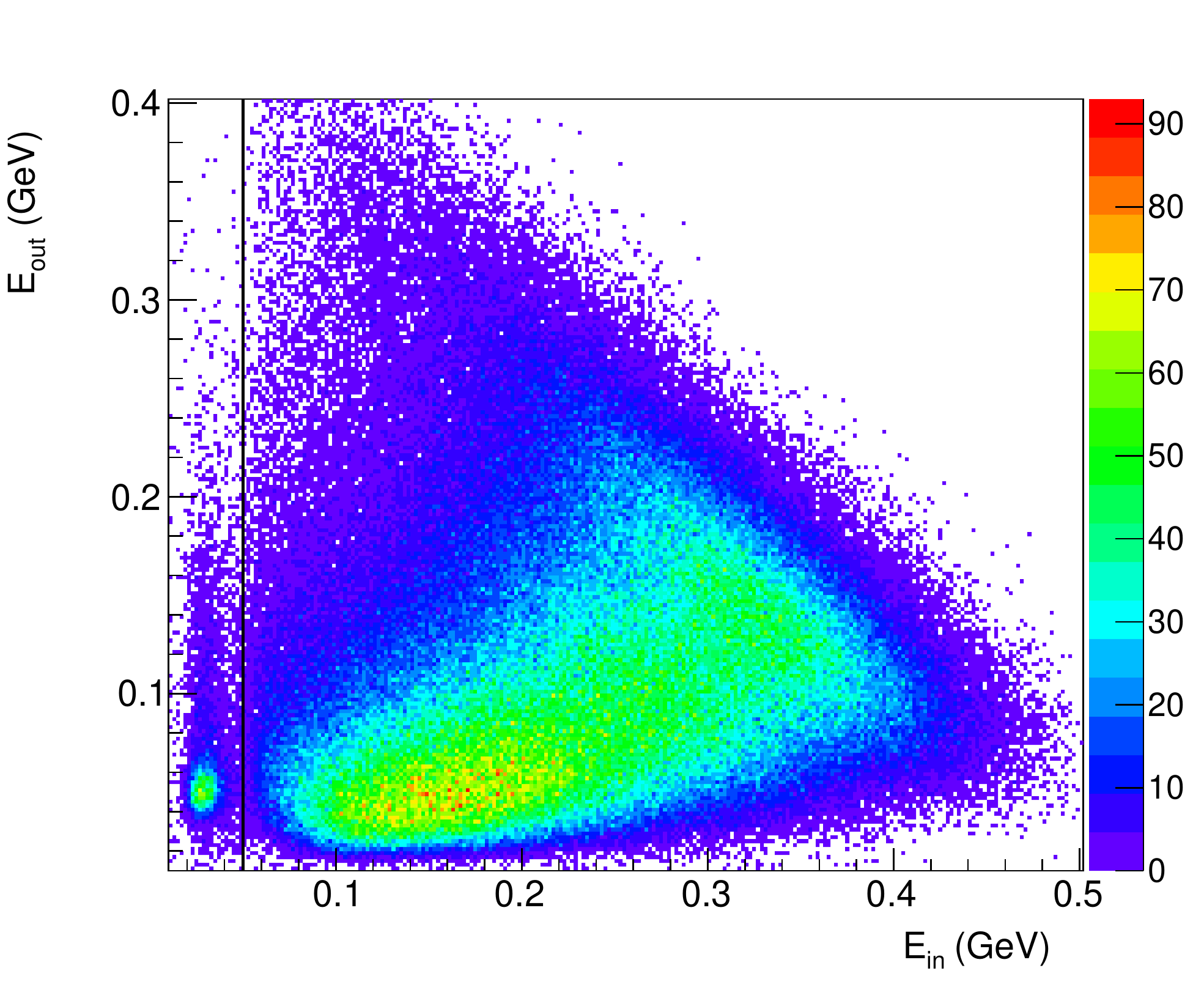, width = 8 cm}
    \caption{(Color online) Energy deposited by negatively charged particles in the inner calorimeter versus energy deposited in the outer calorimeter. Pions are seen at small E$_{in}$ and suppressed with a cut at E$_{in}$ = 50 MeV, represented by the black line. The color ($z$) axis represents the number of events.}
    \label{eIDMinMom}
\end{center}
\end{figure}

\begin{figure}[htb]
  \begin{center}
    \epsfig{file =
   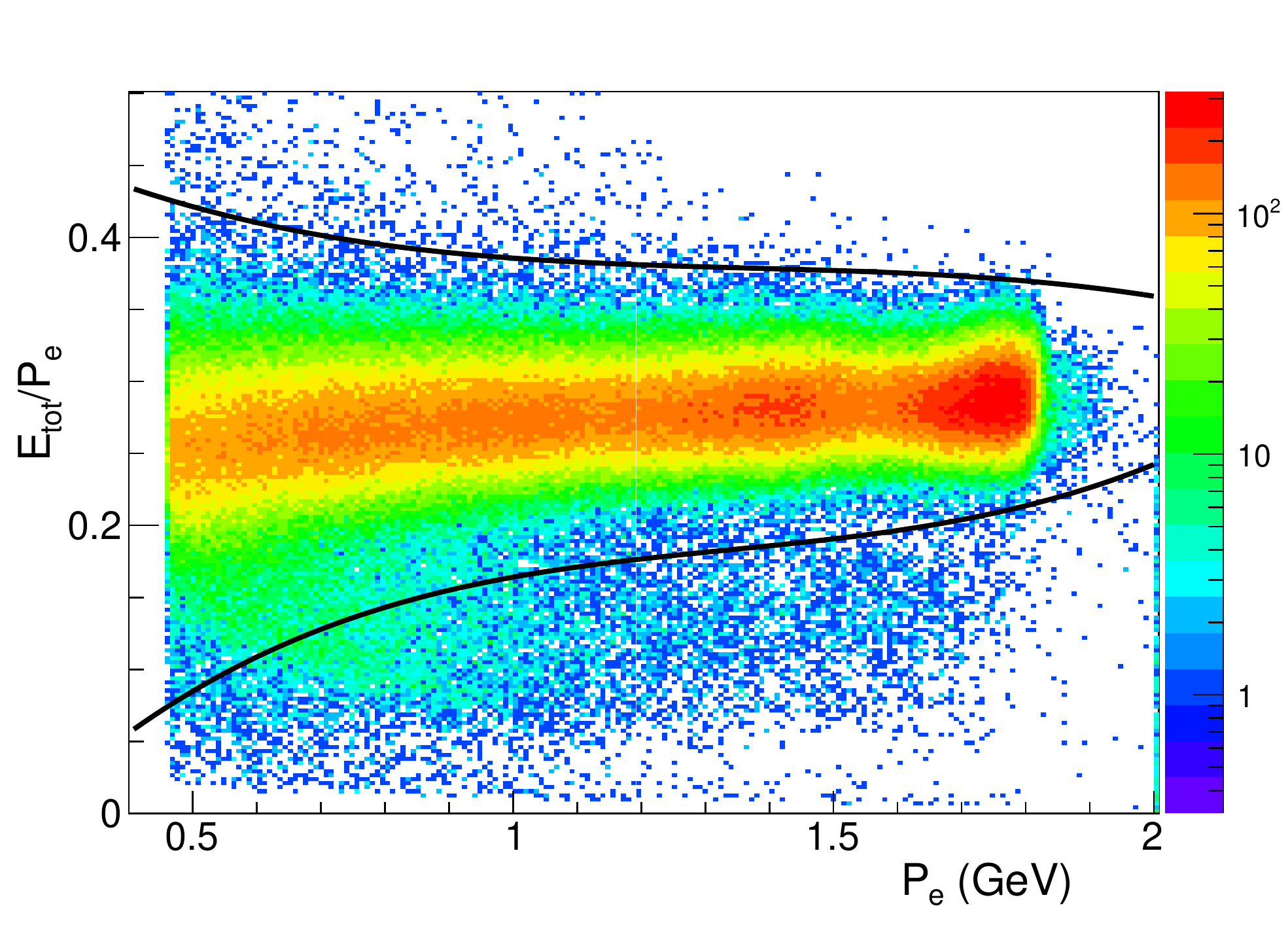, width = 8 cm}
    \caption{(Color online) Energy deposited by negatively charged particles in the calorimeter divided by the momentum of the particles as a function of the momentum. The black curve indicates the 4$\sigma$ cut for selecting electrons. The cut also minimized residual pion contamination below the electron band. The color ($z$) axis represents the number of events.}
    \label{eID}
\end{center}
\end{figure}

The EC is divided into inner and outer modules with independent readout. A 50 MeV threshold on the inner calorimeter is used to reject triggers from hadronic interactions. In the offline analysis, a corresponding cut on the energy deposited in the inner calorimeter suppresses residual pion contamination as shown in Fig.~\ref{eIDMinMom}.
Further electron identification uses the calorimeter energy information along with the particle momentum, reconstructed from charged particle tracking. The ratio of the energy deposited in the EC to the particle momentum as a function of the track momentum is shown in Fig.~\ref{eID} along with our $4\sigma$ electron selection cut. 

\subsection{Proton identification}
\begin{figure}[htb]
  \begin{center}
    \epsfig{file =
   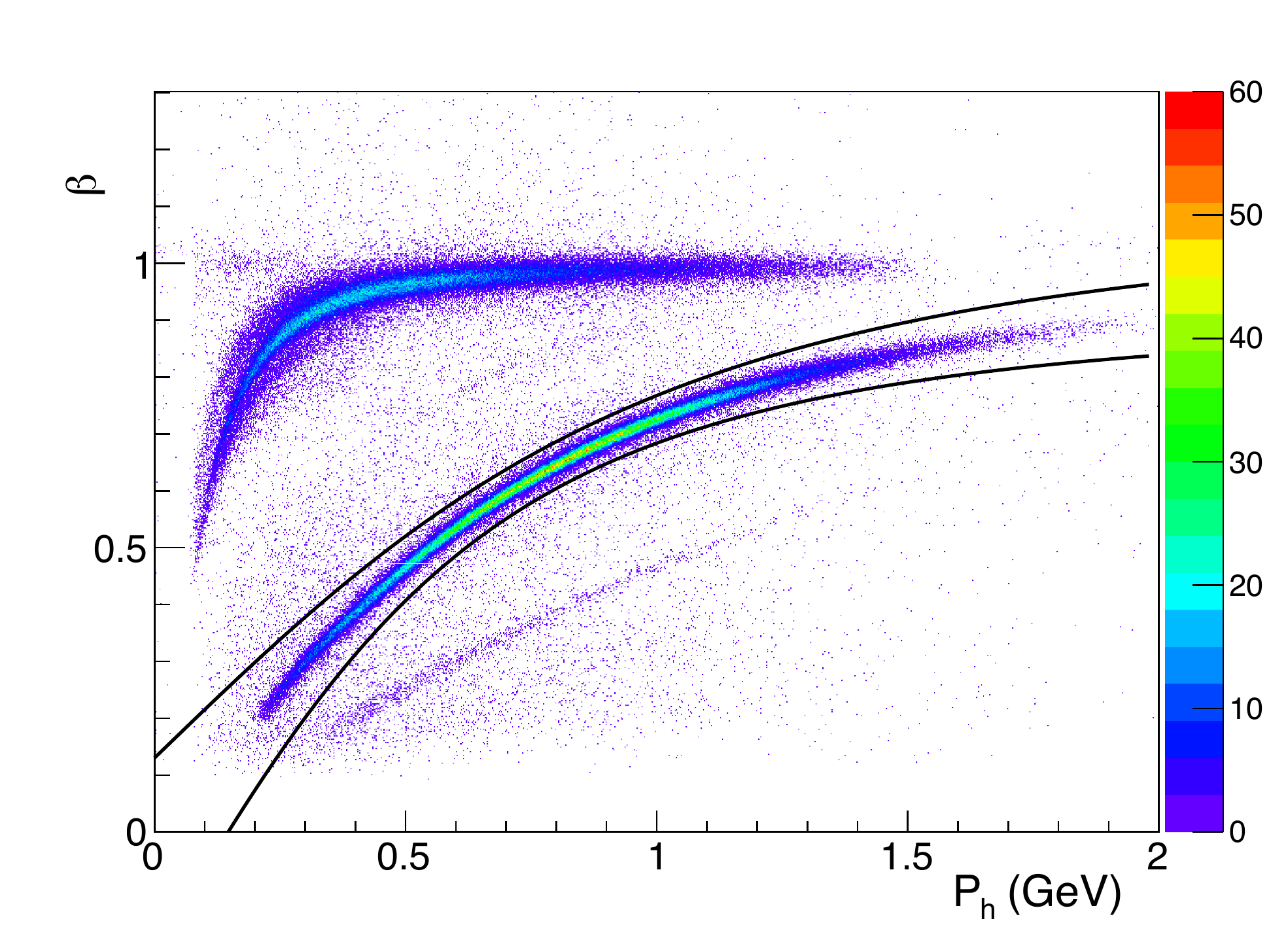, width = 8.5 cm}
    \caption{(Color online) $\beta$ versus momentum for positively charged particles. The solid lines show the cut used to select protons. The bands above the proton band are from $K^{+}$, $\pi^{+}$, and $e^{+}/\mu^{+}$ tracks, while deuterons are visible below the proton band. The color ($z$) axis represents the number of events.}
        \label{pID}
\end{center}
\end{figure}

Proton identification is based on separate measurements of particle velocity and momentum to determine the mass. The velocity $v$, expressed as $\beta=v/c$, is reconstructed from the SC estimate of the track time and the DC estimate of the track length. The distribution of $\beta$ versus momentum for positively charged particles is shown in Fig.~\ref{pID}.  The cut used to select protons is asymmetric with a width of $+4\sigma$, $-5\sigma$, since most of the contamination stemmed from lighter positively charged pions.  

\section{Event selection}
\label{eventSel}
\subsection{Fiducial cuts}
\begin{figure}[htb]
  \begin{center}
    \epsfig{file =
    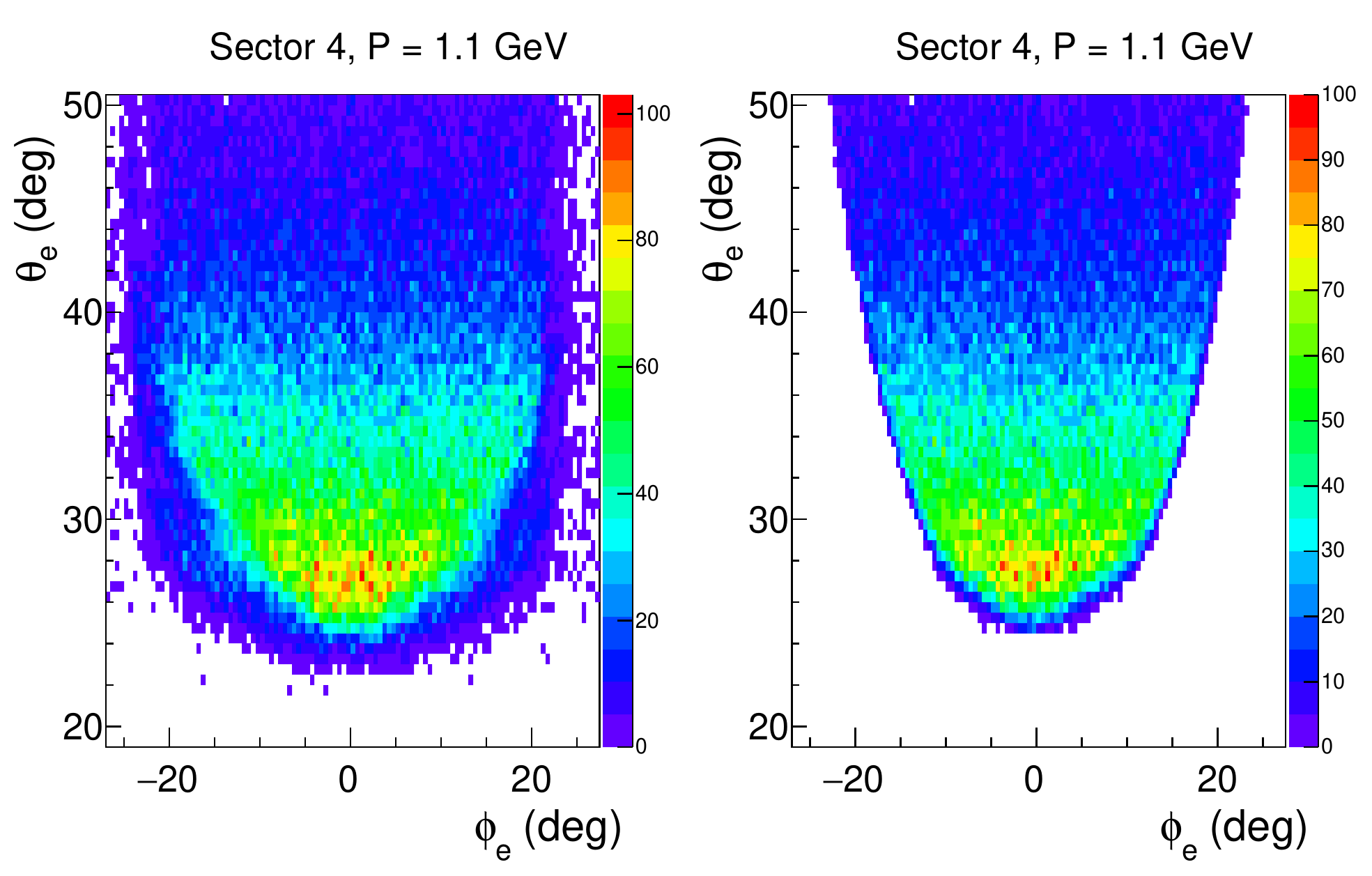, width = 8.5 cm}
    \caption{(Color online) Fiducial region selection for electrons. The angular distributions of events before (left panel) and after (right panel) the fiducial cuts are shown. The regions with low detector efficiency were cut out. The color ($z$) axis represents the number of events.}
        \label{eFid}
\end{center}
\end{figure}

The active area of CLAS is limited by the toroid magnet superconducting coils and the border regions of the detectors. The active area used for data analysis is defined by using fiducial volumes. These volumes are different for protons and electrons and are momentum and sector dependent. An example of a fiducial volume for electrons is shown in Fig.~\ref{eFid}.

\subsection{Target Cuts}
\begin{figure}[htb]
  \begin{center}
    \epsfig{file =
   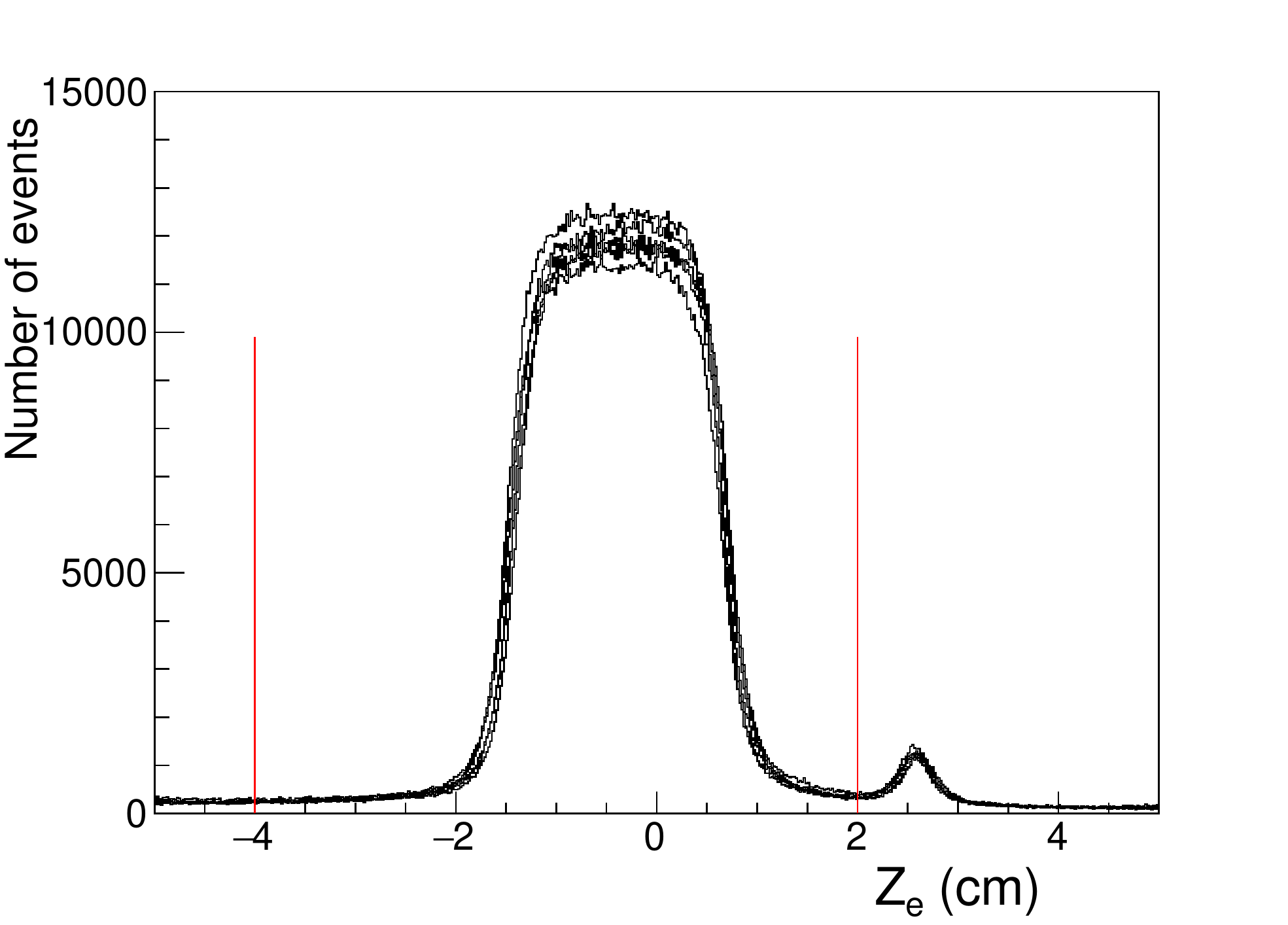, width = 8 cm}
    \caption{(Color online) $Z$ coordinate of the electron vertex for the electrons in different sectors (different curves). The vertex cuts are shown by the red lines.}
    \label{eZElectron}
\end{center}
\end{figure}

The target cell is located near the center of CLAS, shifted upstream by 0.4 cm. Since the target is not centered exactly at (0, 0) in the ($x$, $y$) coordinates transverse to the beam line, the reconstructed position of the reaction vertex deviates from the actual position, requiring a sector-dependent correction.  The correction is based on the DC geometry and uses the fact that if the beam is not centered at (0, 0), the reconstructed $z$ position will have a sin$\phi$ modulation. The actual average beam position is at (0.187 cm, -0.208 cm) and this value is used to align the $z$ position of the vertex.  A cut is made to select events originating from the target (see Fig.~\ref{eZElectron}). The same correction was later applied to protons and a cut on the difference between the vertex position of the proton and electron was applied. We used the same beam position of  (0.187 cm, -0.208 cm) in the simulation and applied exactly the same correction and cuts.
 
\subsection{Channel identification}
Although it is possible to identify a $\pi^{0}$ in CLAS from the $\pi^0\rightarrow 2\gamma$ decay by reconstructing the invariant mass of two photons in the calorimeters, the limited acceptance will impose unnecessary limitations on the statistical precision. Instead, we can reconstruct the four-vector of the missing particle $X$ in the $ep \to e'p' X$ reaction using  the initial and scattered four-momenta of the electron and proton along with energy and momentum conservation. For exclusive $e'p'\pi^{0}$ events, the $m_{X}$ distribution should show a peak at the mass of the $\pi^{0}$.
\begin{figure}[ht]
  \begin{center}
    \epsfig{file =
   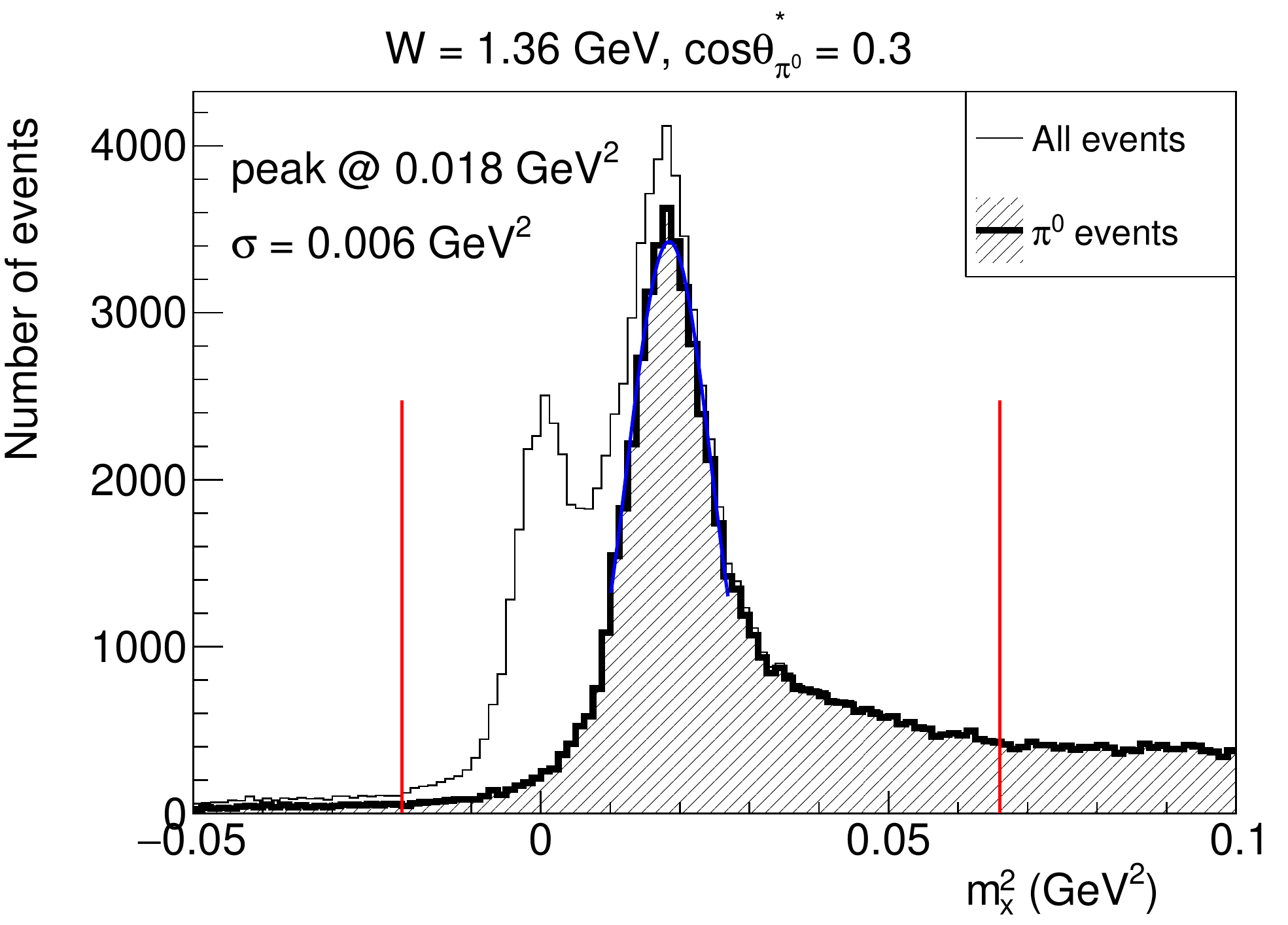, width =8.5cm}
  \caption{(Color online) Bethe-Heitler (BH) event separation. One cannot reliably separate BH events (peak around zero) from $\pi^{0}$ events (peak around 0.02~GeV$^{2}$) using only a missing mass cut. A more sophisticated procedure, based on the reaction kinematics is needed to provide the $\pi^{0}$ event distribution (shaded area). Blue line is the gaussian fit to the peak. The red lines are the final exclusivity cuts.}
    \label{nEv4D}
  \end{center}
\end{figure}

\begin{figure}[ht]
  \begin{center}
    \epsfig{file =
   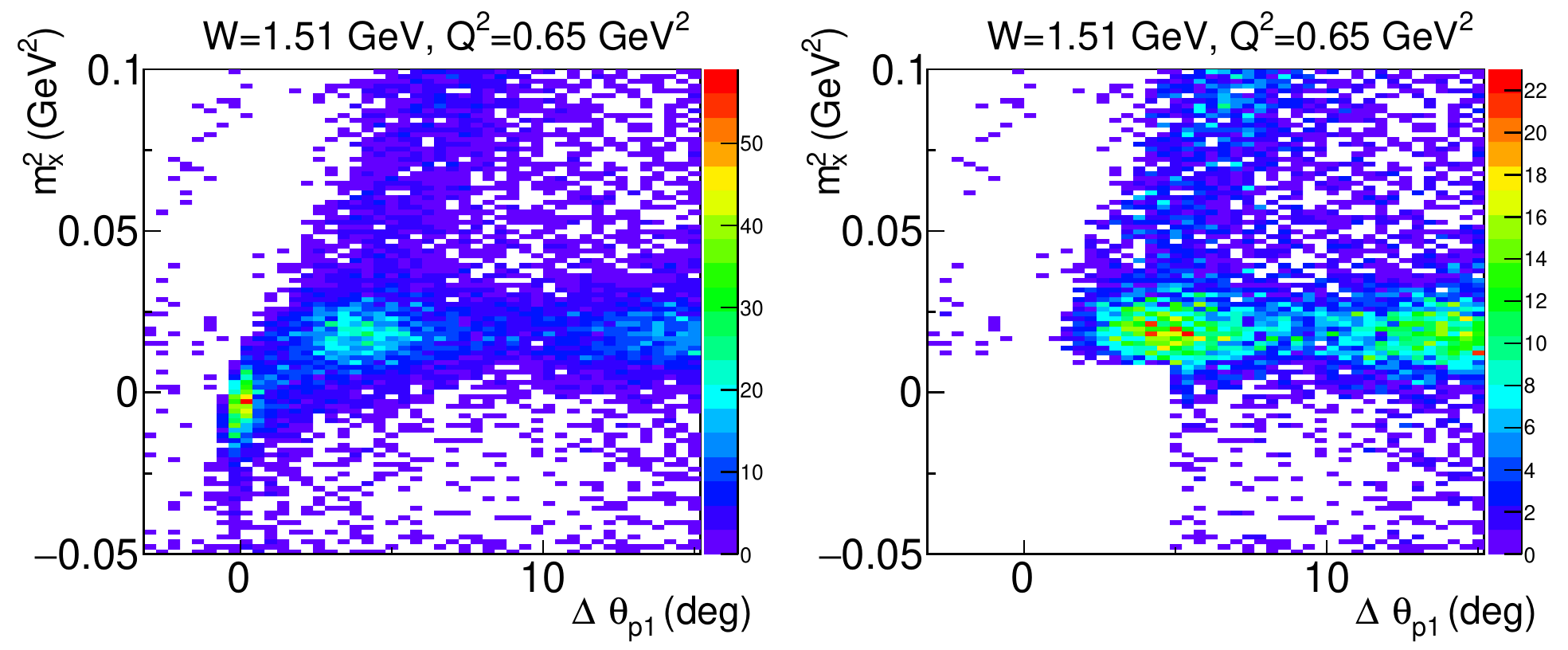, width =9cm}
  \caption{(Color online) Bethe-Heitler (BH) event separation using post-radiative  kinematics. Post-radiative events are concentrated in the $\Delta\theta_{p1} = 0^{o}$, $m_{X}^{2} = 0$~GeV$^{2}$ region on the left plot, where no BH separation cuts were applied. The sample of the clean $\pi^{0}$ events is presented on the right plot, where all the BH separation cuts were applied. The color ($z$) axis represents the number of events.}
    \label{deltaTheta1Theta2CutPionID}
  \end{center}
\end{figure}

The overlap of the elastic and elastic radiative events, which constitutes the majority of the background, with the single pion events in the missing mass squared spectrum (see Fig.~\ref{nEv4D}) does not allow for a complete  separation using only a simple missing mass cut. Instead, the choice of a suitable topology allows for the separation of exclusive single $\pi^{0}$ events from the Bethe-Heitler (BH) background. We use three cuts on different variables to perform the event separation: (1) Center-of Mass pion angle $\phi_{\pi^{0}}$ as a function of the missing mass squared, the difference between the measured and reconstructed polar angle of the proton $\theta_{p}$ in the assumption of the (2) post-radiative $\theta_{p_{1}}$ (see Eq.~\ref{theta1Eq}) and (3) pre-radiative BH events $\theta_{p_{2}}$ (see Eq.~\ref{theta2Eq}). In case of the first distribution, the BH events concentrate around $\phi_{\pi^{0}}$ = 0, while the exclusive $\pi^{0}$ events are distributed uniformly. In case of the second and third distributions, the difference between the measured and reconstructed proton $\theta_{p}$, post- and pre-radiative events also concentrate around 0 for the BH events in the corresponding kinematics (Fig.~\ref{deltaTheta1Theta2CutPionID} represents the post-radiative kinematics). This allows for reliable $\pi^{0}$ separation.  The resulting missing mass squared distribution is shown in Fig.~\ref{nEv4D}.
\begin{equation}
tan\theta_{1} = \frac{1}{(1 +
  \frac{E}{M_{p}})tan\frac{\theta_{e'}}{2}}
\label{theta1Eq}
\end{equation}

\begin{equation}
tan\theta_{2} = \frac{1}{(1 +
  \frac{E_{f}}{M_{p} - E_{f} + E_{f}cos\theta_{e'}})tan\frac{\theta_{e'}}{2}}
\label{theta2Eq}
\end{equation}

A cut on the upper value of $m_{X}^{2} < 0.066$~GeV$^{2}$ is necessary in order to limit the contribution of radiative $\pi^{0}$ events. This cut is accounted for in both simulation and in the calculations of the radiative corrections. The last cut on the lower value of the $m_{X}^{2} > -0.02$~GeV$^{2}$ finalizes our exclusive event selection.
\subsection{Kinematic binning}
\begin{figure}[htp]
\begin{center}
\includegraphics[width=8cm]{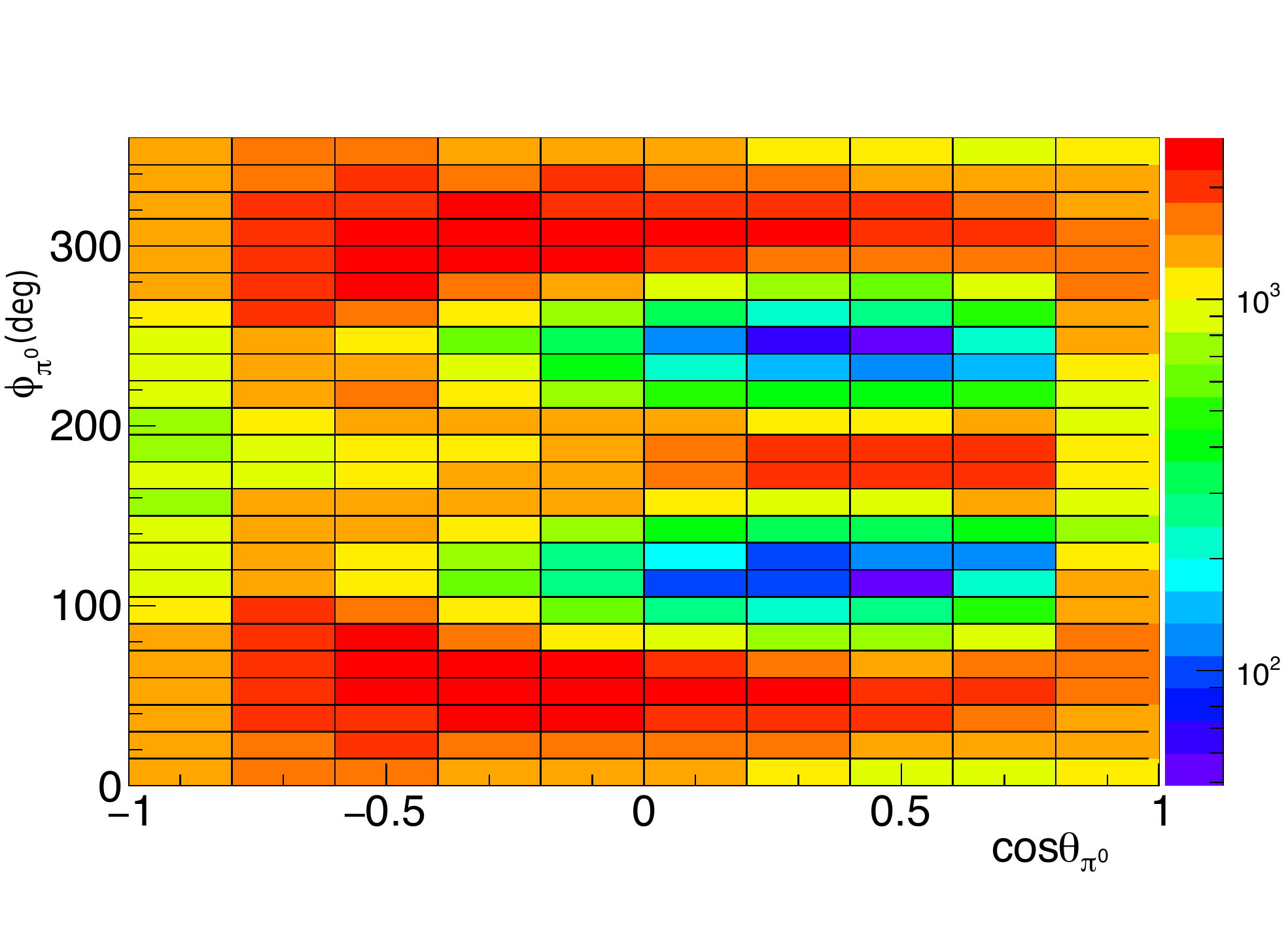}
\caption{(Color online) Coverage and binning in cos$\theta_{\pi^{0}}$ and $\phi_{\pi^{0}}$  (indicated by black lines)
  for the $\pi^{0}$ electroproduction events, before acceptance corrections.}
\label{cosThetaPhiBinning}
\end{center}
\end{figure}

The $ep \to e'p'\pi^{0}$ kinematics is defined by four variables: $W$, $Q^{2}$, cos$\theta_{\pi^{0}}$, and  $\phi_{\pi^{0}}$. Bins in $W$ were chosen to observe cross section variations due to contributions from individual resonances, while the $Q^{2}$ binning was optimized to cover the rapid cross section variation with the increase of photon virtuality. Since the extraction of the structure functions was performed by fitting the cross section over $\phi_{\pi^{0}}$, the bin size was chosen to adequately sample the variations of the CLAS acceptance over this variable to minimize systematic uncertainties in the acceptance corrections. This dataset covered a wide $W$ and $Q^{2}$ range (see Fig.~\ref{wQ2Binning} and Table~\ref{WQ2BinningTable}) and the CLAS acceptance allowed coverage over nearly the full angular range in the center-of-mass system (see Fig.~\ref{cosThetaPhiBinning} and Table~\ref{cosThetaPhiBinningTable}).
\begin{table}
\centering
\begin{tabular}{|c|c|c|c|c|}
\hline Variable& Bin size& Number of bins&Lower limit&Upper limit\\
\hline $W$, GeV & 0.025& 28& 1.1& 1.8\\
\hline $Q^{2}$, GeV$^{2}$ & 0.1 & 6& 0.4& 1.0\\
\hline
\end{tabular}
\caption{$W$ and $Q^{2}$ binning of the experiment.}
\label{WQ2BinningTable}
\end{table}

\begin{table}
\centering
\begin{tabular}{|c|c|c|c|c|}
\hline Variable& Bin size& Number of bins&Lower limit&Upper  limit\\
\hline cos$\theta_{\pi^{0}}$ & 0.2& 10& -1& 1\\
\hline $\phi_{\pi^{0}}$ & 15$^{\circ}$ & 24& 0$^{\circ}$& 360$^{\circ}$\\
\hline
\end{tabular}
\caption{Binning in cos$\theta_{\pi^{0}}$ and $\phi_{\pi^{0}}$.}
\label{cosThetaPhiBinningTable}
\end{table}

\begin{figure}[htp]
\begin{center}
\includegraphics[width=8cm]{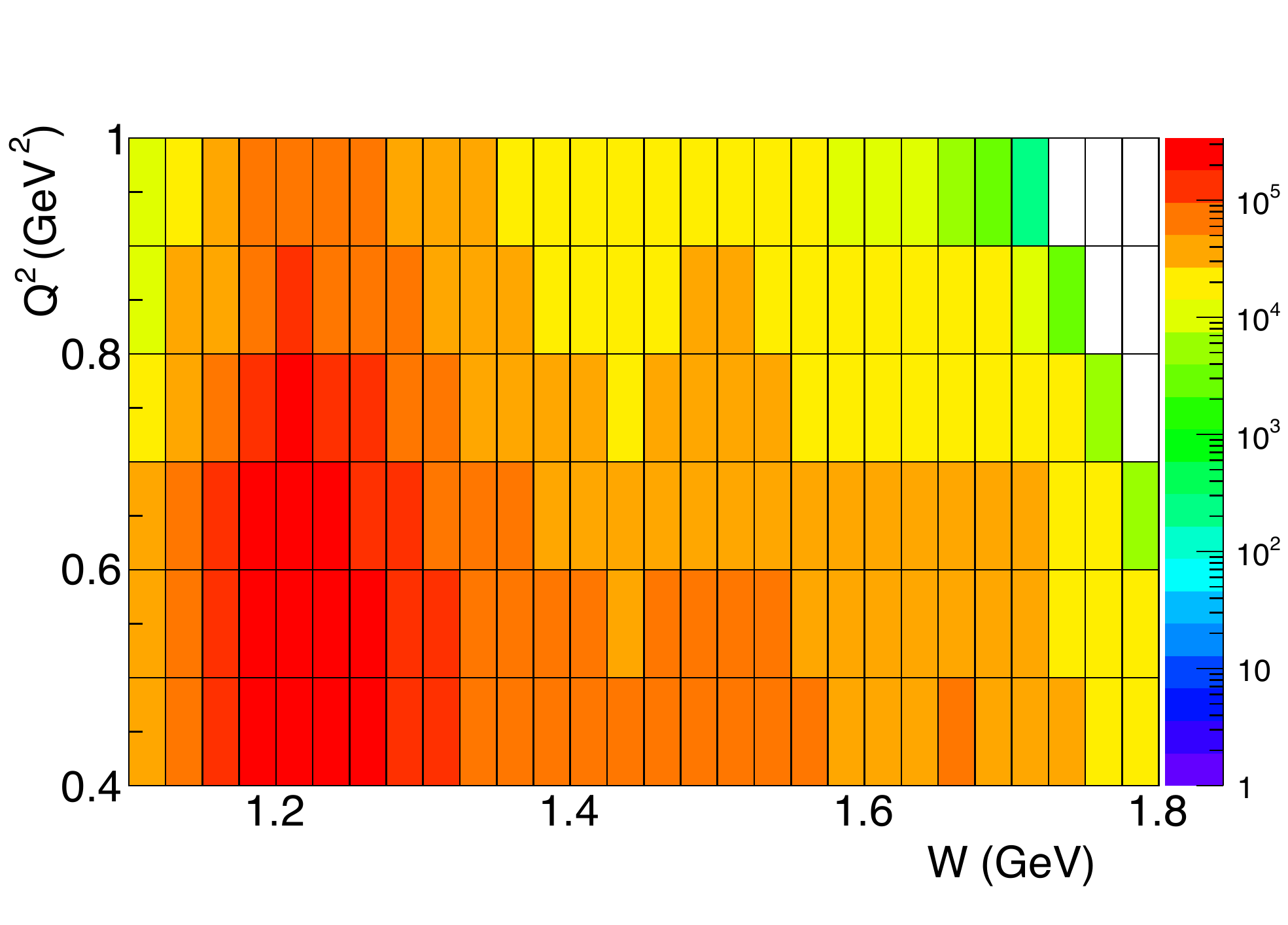}
\caption{(Color online) Coverage and binning in $W$ and $Q^{2}$ (indicated by black lines) for the $\pi^{0}$ electroproduction
  events, before acceptance corrections.}
\label{wQ2Binning}
\label{wQ2Binning}
\end{center}
\end{figure}

\section{Normalization}
\subsection{Dataset selection}
Conditions during data taking can vary, for instance due to target density fluctuations, beam quality, or conditions on the data acquisition. However, the exclusive $\pi^{0}$ event yield, normalized to the total accumulated charge measured by Faraday Cup, should be a constant.  The distribution of normalized yields over time was fitted with a Gaussian and acceptable conditions were defined by requiring the normalized yield to be within $\pm 3\sigma$ of the mean.
\subsection{Elastic cross section}
\begin{figure*}[htp]
\begin{center}
\includegraphics[width=10cm]{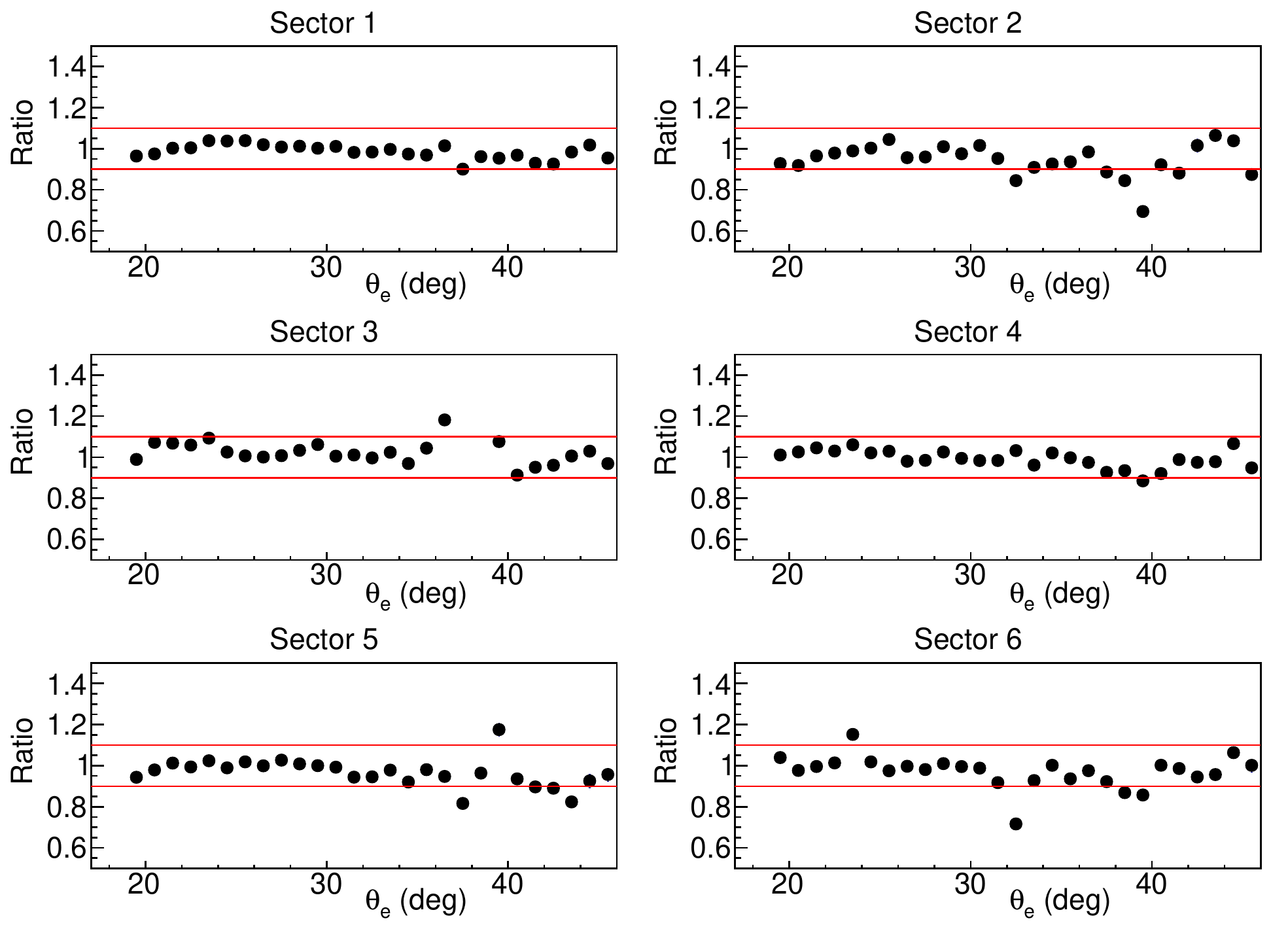}
\caption{(Color online)  Ratio of the elastic cross section with detection of the electron and proton, measured experimentally, compared to the Bosted~\cite{BostedParametrization}  parameterization. Statistical error bars are within the marker size. The red lines are at $\pm10 \%$ about unity. The agreement between the data and model is well within $10\%$ on average.}
\label{csRatio}
\end{center}
\end{figure*}

Using a well known benchmark reaction one can independently cross check procedures used to obtain the final results.  In this work, the exclusive $ep$ elastic cross section was measured simultaneously with the inelastic data, to monitor the Faraday Cup performance and the detector calibrations, as well as the electron and proton identification procedures and fiducial cuts. Procedure, similar to one used in the~\cite{bedlinskNorm} is used to estimate the $E_{TOF}$, which is found to be of the order of 5\%.  The experimentally measured cross sections normalized to a parameterization of Bosted~\cite{BostedParametrization} is plotted in Fig.~\ref{csRatio} for each CLAS sector as a function of the scattered electron angle. 
\section{Corrections} 
\label{corrections}
\subsection{Target wall subtraction}
Exclusive $\pi^{0}$ events can originate both from within the $LH_{2}$ target volume and from the upstream/downstream windows of the target cell. These windows are made of 15~$\mu m$ aluminum foil. Since our vertex resolution combined with the short target length does not permit a vertex cut, empty target runs were used to estimate the background yields. To make a proper correction, exactly the same particle identification procedure, including electron, proton, and $\pi^{0}$ identification, is applied to the empty target run dataset. Subsequently, these events are divided into the same ($W, Q^{2},  \mathrm{cos} \theta_{\pi^{0}}, \phi_{\pi^{0}}$) bins as the full target events (see Tables~\ref{WQ2BinningTable} and~\ref{cosThetaPhiBinningTable}), normalized by the corresponding Faraday cup charge, and subtracted from the final sample. The average value of the correction over the whole phase space is less then 5\%.
\subsection{Acceptance corrections}
There are two major factors that determine the detector acceptance: geometrical acceptance, which limits the area in which particles could possibly be detected, and detector efficiency. Both are accounted for using GSIM~\cite{GSIM}, a GEANT-based simulation of the CLAS detector, which includes the actual detector geometry and materials. Magnetic field maps used in the simulation are results of the Finite Element Analysis calculations. Certain detector inefficiencies, including dead wires in the drift chambers and missing channels in the photomultiplier tube (PMT) based detectors, are incorporated as well.  

The detector acceptance is defined as 
\begin{equation}
A(W, Q^{2}, \mathrm{cos}\theta_{\pi^{0}}, \phi_{\pi^{0}}) = \frac{N_{rec}(W, Q^{2}, \mathrm{cos}\theta_{\pi^{0}}, \phi_{\pi^{0}})} {N_{gen}(W, Q^{2}, \mathrm{cos}\theta_{\pi^{0}}, \phi_{\pi^{0}})},
\label{acceptanceEquation}
\end{equation}
where $N_{rec}$ and $N_{gen}$ are the number of reconstructed and generated $ep \to e'p'\pi^{0}$ Monte Carlo events, respectively, for a given kinematical bin. The event generator was based on the convolution of the MAID07~\cite{MAID} unitary isobar model with a Mo-Tsai~\cite{MoTsai} radiation model. The output of the GSIM code was then reconstructed in the same way as the experimental data from the detector. 

\begin{figure*}[htp]
\begin{center}
\includegraphics[width=12cm]{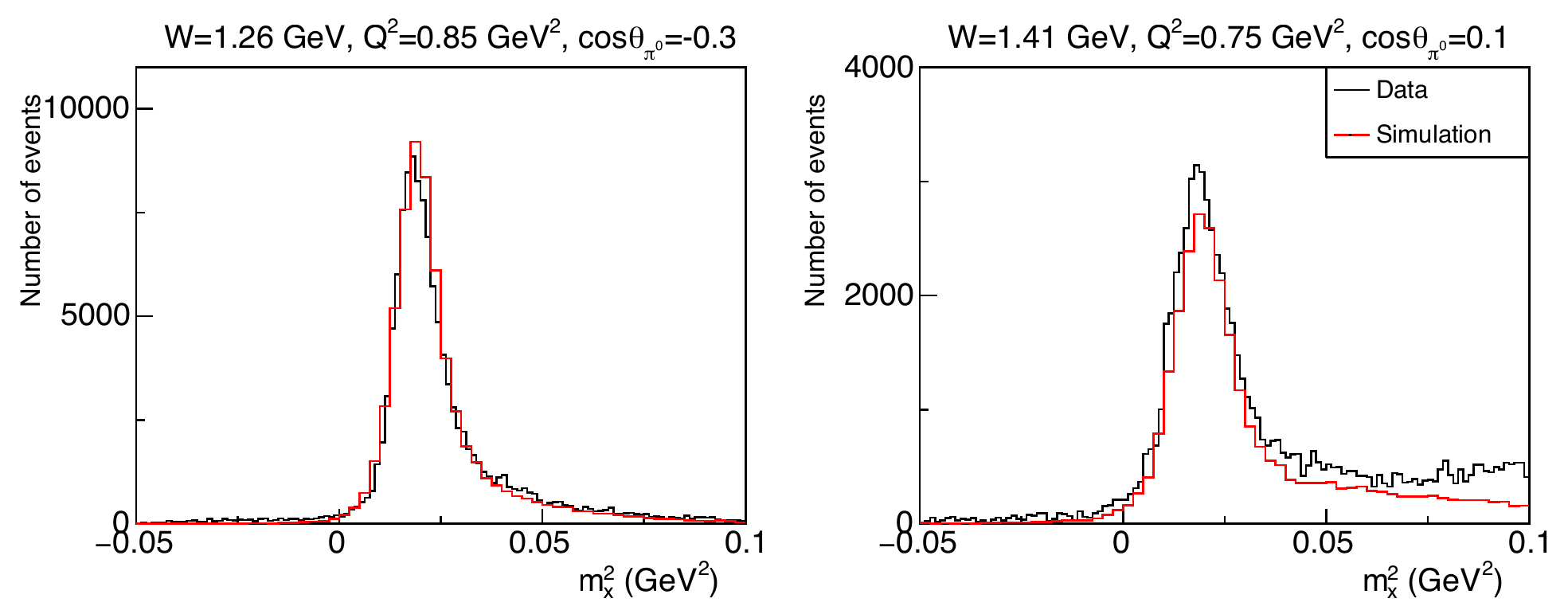}
\caption{(Color online) Missing mass squared distribution for data (black lines) and  simulation (red lines) overlapped, plotted for different representative $W$, $Q^{2}$ and cos$\theta_{\pi^{0}}$ values, covering a wide range of kinematics. The normalization factor was chosen as the ratio of the  total number of the $\pi^{0}$ events in data and simulation and is the same for all panels.}
\label{missingMassPeakDataSimLowW}
\end{center}
\end{figure*}

\begin{figure}[htp]
\begin{center}
\includegraphics[width=8cm]{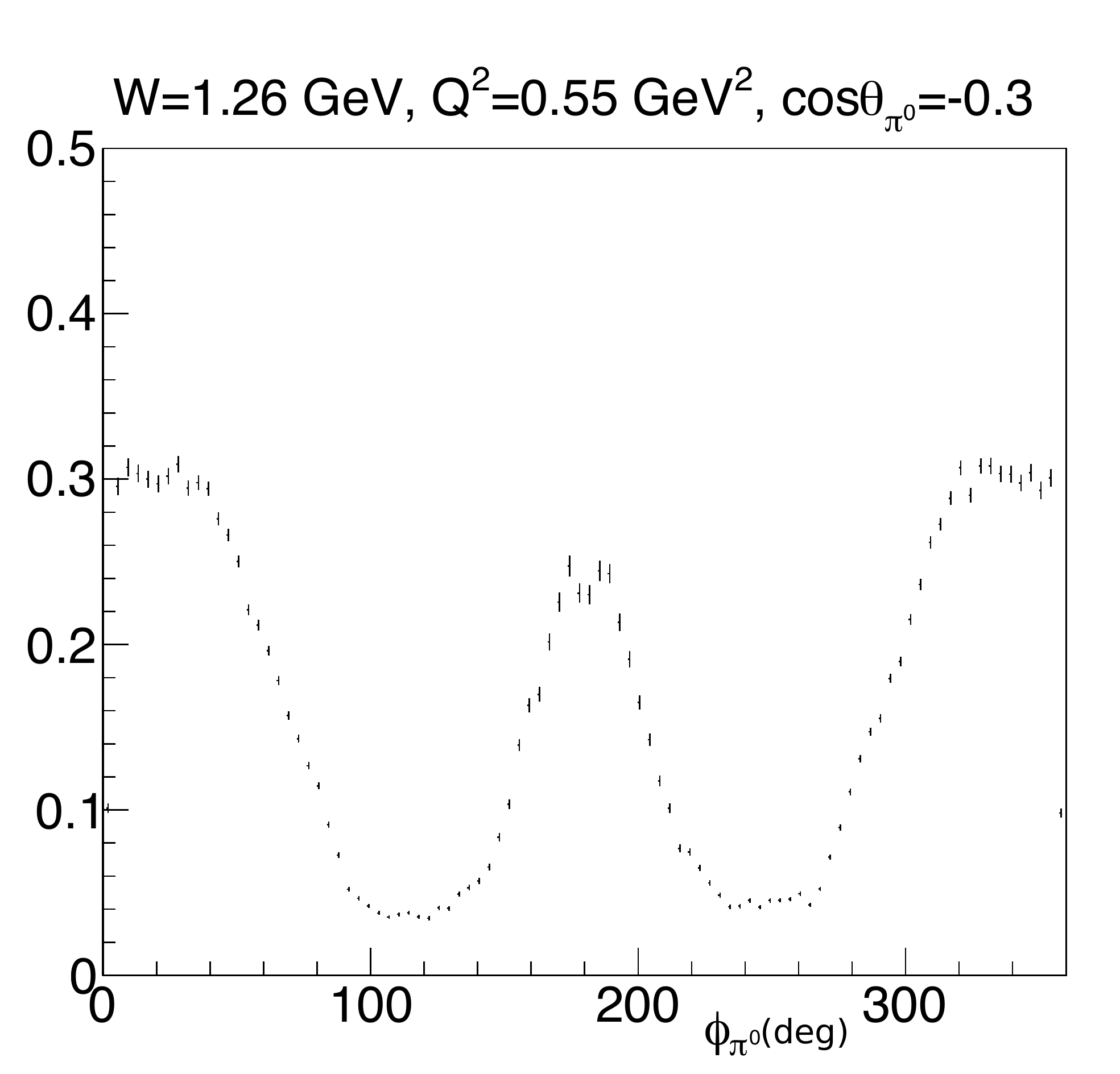}
\caption{Acceptance correction as a function of
 $\phi_{\pi^{0}}$ for $W =
  1.2625$~GeV, $Q^{2}$ = 0.55~GeV$^{2}$,  cos$\theta_{\pi^{0}} = -0.3$.}
\label{accSample}
\end{center}
\end{figure}

Reconstructed events have to closely follow the energy and angular resolution of the actual CLAS data so that one could apply the same event selection criteria for both data and simulation. The comparison of both for the $e'p'$ missing mass squared is shown in Fig.~\ref{missingMassPeakDataSimLowW} and serves as an illustration of the good agreement between data and simulation over a wide kinematical ranges.
A sample acceptance distribution is presented in Fig.~\ref{accSample} for a single kinematic bin.
\subsection{Radiative corrections}
\begin{figure*}[htp]
\begin{center}
\includegraphics[width=16cm]{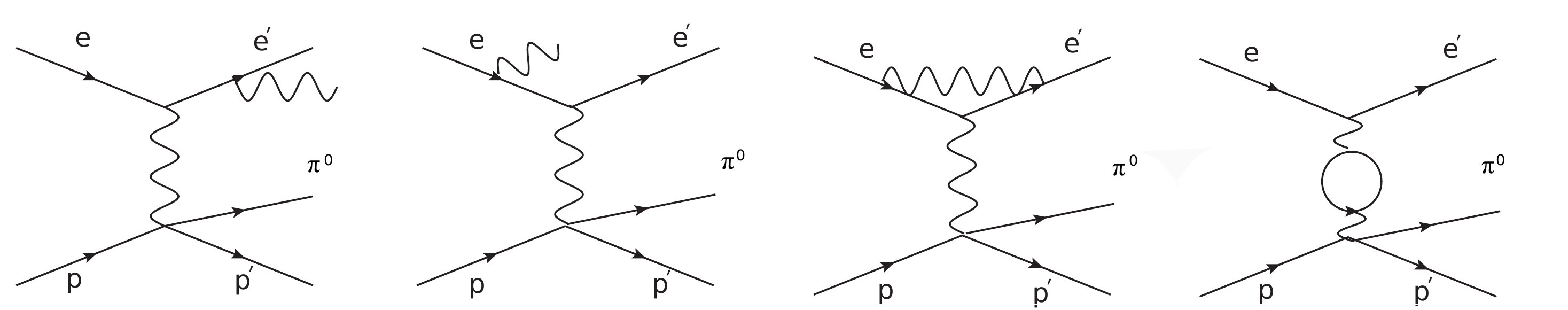}
\caption{Left to right: Post radiative bremsstrahlung radiation, pre-radiative bremsstrahlung radiation, vertex modification, and vacuum polarization.}
\label{pictureRadiativeProcesses}
\end{center}
\end{figure*}

\begin{figure}[htp]
\begin{center}
\includegraphics[width=8cm]{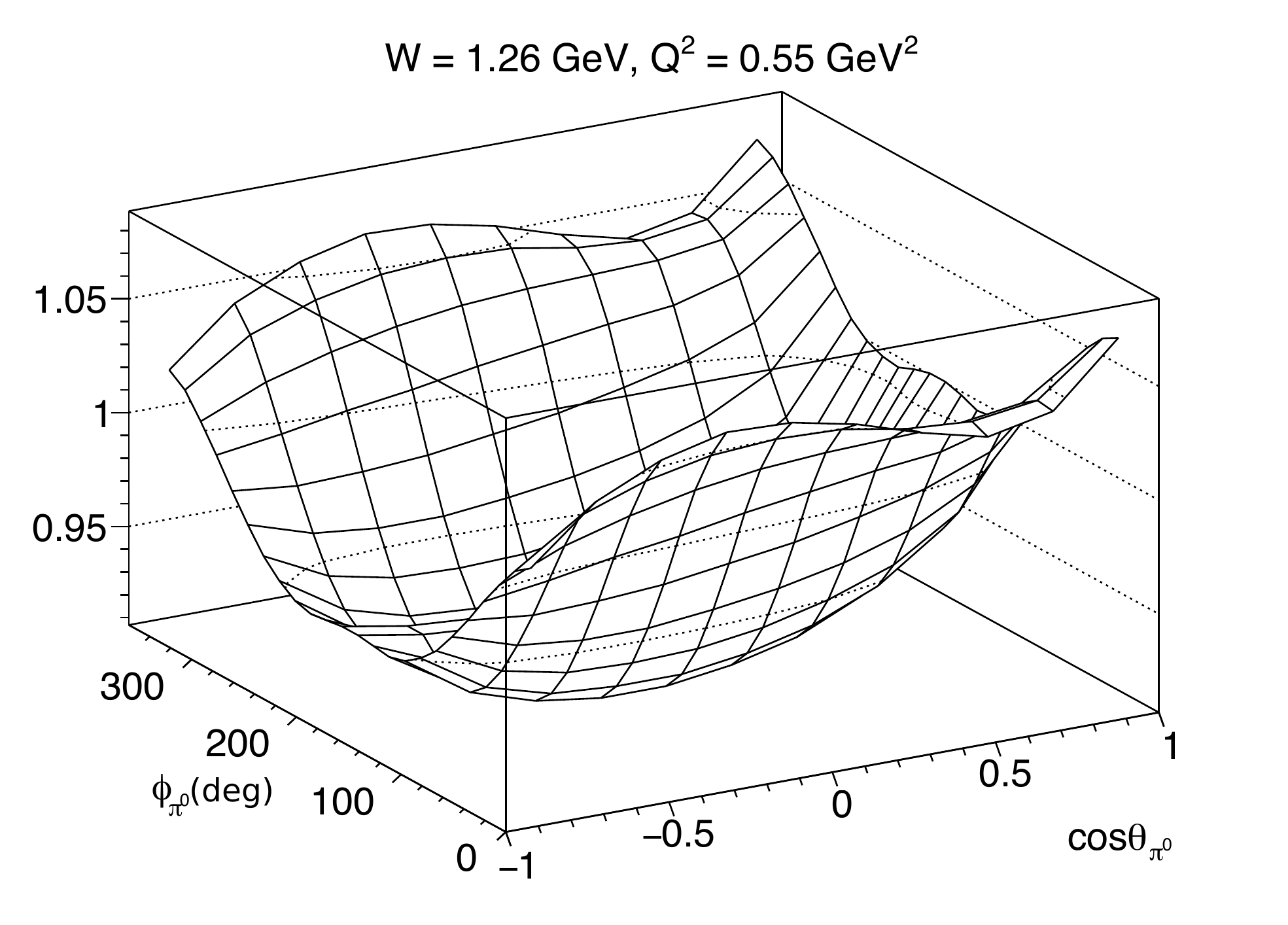}
\caption{Radiative correction as a function of
 $\phi_{\pi^{0}}$ and $\cos\theta_{\pi^{0}}$ for $W =
  1.2625~$GeV, $Q^{2} = 0.55~$GeV$^{2}$.}
\label{radCorr2D}
\end{center}
\end{figure}

Internal bremsstrahlung diagrams  such as presented in Fig.~\ref{pictureRadiativeProcesses} distort the experimentally measured cross sections. These distortions were calculated exactly for single pion electroproduction off the proton using the EXCLURAD approach developed in~\cite{exclurad}. The corrections require a model cross section that accounts for all four structure functions. A multiplicative correction can then be obtained by dividing the radiated model cross section by the unradiated model: 
\begin{equation}
R(W, Q^{2},  \mathrm{cos}\theta_{\pi^{0}}, \phi_{\pi^{0}}) = \frac{\sigma_{RAD}(W, Q^{2},  \mathrm{cos}\theta_{\pi^{0}}, \phi_{\pi^{0}})} {\sigma_{NORAD}(W, Q^{2},  \mathrm{cos}\theta_{\pi^{0}}, \phi_{\pi^{0}})}.
\label{rcEquation}
\end{equation}
The MAID07 predictions were used as the model input.
To account for possible variations of the radiative correction inside the bin, all bins were subdivided into three smaller bins over each of four kinematical variables ($W, Q^{2}$,  cos$\theta_{\pi^{0}}, \phi_{\pi^{0}}$). Radiative corrections were then calculated independently in each of 81 (3$^{4}$) of the smaller bins, and the average over these 81 bins was used for the final corrections. An example of the center-of-mass angular dependence of the corrections for one ($W, Q^{2}$) bin is presented in Fig.~\ref{radCorr2D}.

\subsection{Bin centering corrections}
The cross section might not vary linearly across the width of a bin, which would result in the calculated cross section at the bin center not coinciding with the average value of the cross section in that bin.  MAID07 was used to evaluate the corrections. We  divided each bin over ($W, Q^{2}$,  $\cos\theta_{\pi^{0}}, \phi_{\pi^{0}})$ into ten smaller bins, calculated the cross section in the center of each of the smaller bins ({$CS_{av}$), and separately calculated the cross section in the center of the large bin ($CS_{c}$). The bin centering correction was then defined as
\begin{equation}
B(W, Q^{2},  \mathrm{cos}\theta_{\pi^{0}}, \phi_{\pi^{0}}) = \frac{CS_{av}}{CS_{c}},
\label{bcEquation}
\end{equation}
with the example for a single kinematic bin shown in Fig.~\ref{bcCorr1D}.
\begin{figure}[htp]
\begin{center}
\includegraphics[width=8cm]{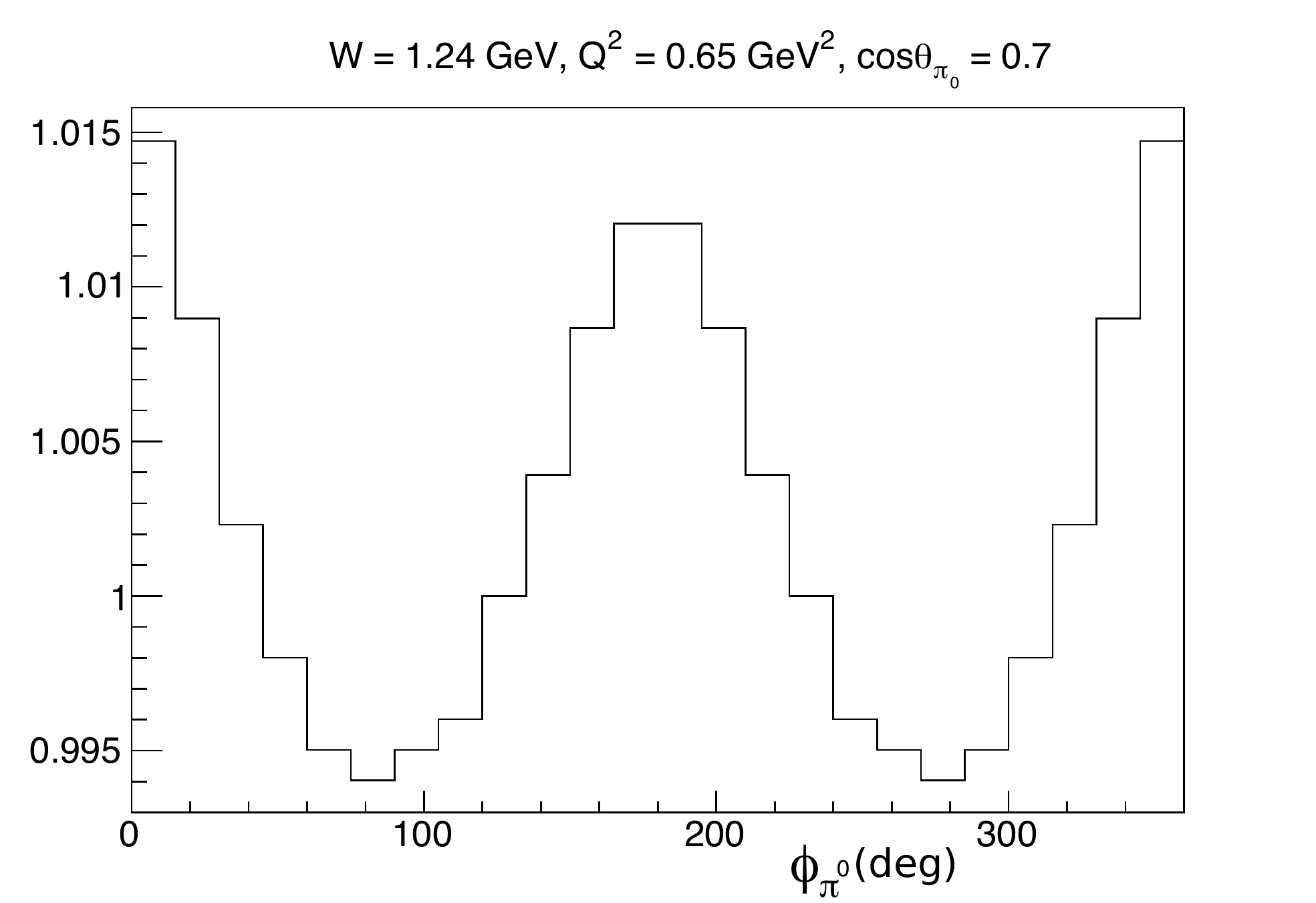}
\caption{Bin centering correction as a function of
 $\phi_{\pi^{0}}$ for $W =
  1.2375~$GeV, $Q^{2} = 0.65~$GeV$^{2}$,  cos$\theta_{\pi^{0}} = 0.7$.}
\label{bcCorr1D}
\end{center}
\end{figure}
\section{Systematic uncertainties}
The high statistical precision of these data required an extensive study of possible sources of systematic uncertainties in order to characterize the reliability of the results. The general method of the uncertainty calculation was to vary characteristic parameters corresponding to each step in the analysis procedure to quantify the effect on the resulting cross sections and structure functions on a bin-by-bin basis.  The summary of the systematics study is shown in Table~\ref{finalSystematicsTable}, and the overall value of the uncertainty averaged over all kinematical bins, defined as a sum in quadrature of the individual contributions, is equal to 8.7\%.

The most important sources of systematic uncertainties are the fiducial cuts for both electrons and protons, the missing mass cut, and the absolute normalization, which itself served as an integral measure of the quality of electron and proton identification. 
The position of the missing mass cut affected the value of the radiative correction, so for each modification of the cut, the correction was recalculated and included in the reported results.

\begin{table}[htp]
\begin{center}
\begin{tabular}{|l | c | c |}
\hline
Cut & Uncertainty\\
\hline
Sampling fraction & 1.49\%\\
\hline
Electron fiducial cut &  3.80\%\\
\hline
Proton identification &  2.44\%\\
\hline
Proton fiducial cut & 4.1\%\\
\hline
$m_{X}^{2}$ cut & 2.56\%\\
\hline
$\Delta \theta_{1}$ cut & 0.68\%\\
\hline
$\Delta \theta_{2}$ cut & 0.77\%\\
\hline
$\phi_{\pi^{0}}$ cut & 1.92\%\\
\hline
Normalization & 5\%\\\hhline{|=|=|}
Total & 8.7\%\\
\hline
\end{tabular}
\caption{Overview of sources and values of the systematic uncertainties. See text for explanation.}
\label{finalSystematicsTable}
\end{center}
\end{table}
\subsection{Normalization}
The design of CLAS permitted the simultaneous measurement of elastic ($ep\to e'p'$) and inclusive cross sections ($ep\to e'X$) along with the exclusive $\pi^0$ data.  This allowed for a comprehensive check of the electron and proton identification, tracking efficiency, and absolute luminosity, including the Faraday Cup calibration and understanding of the target properties, over the full $W$ range of the exclusive measurement. It also served as a confirmation of the correctness of our simulation procedure, since the detector simulation and event reconstruction are independent of the reaction channel and event generator used.

The elastic cross section, for which both electron and proton were detected, was compared to a parametrization of the available world data~\cite{BostedParametrization} and found to be consistent within 5\%. The inclusive cross section, covering the whole $W$ and $Q^{2}$ range was compared to both the Keppel \cite{keppelInclusive} and Brasse~\cite{brasseInclusive} parameterizations, and display a good agreement in the full kinematical region. 
From this comparison we estimated the normalization uncertainty to be also at the level of 5\%. This value was added to the overall systematic uncertainty.

\section{Results and discussion}
\subsection{Differential cross section}
The cross section obtained from the number of the events $N_{events}$ in the four-dimensional
($W,Q^{2}$, cos~$\theta_{\pi^{0}},\phi_{\pi^{0}}$) bins is given by the expression
\begin{widetext}
\begin{equation}
\frac{d\sigma}{d\Omega_{\pi^{o}}dWdQ^{2}} = N_{events}\frac{1}{N_{e}N_{p}}\frac{1}{R}\frac{1}{AE_{TOF}}B\frac{1}{\Delta W\Delta Q^{2}\Delta \mathrm{cos}\theta_{\pi^{o}}\Delta \phi_{\pi^{o}}}\frac{1}{\Gamma_{v}},
\label{nEvToCS}
\end{equation}
\end{widetext}
where 
\begin{equation}
N_{e} = \frac{Q_{F}}{e}
\end{equation}
 is  the number of electrons delivered to the target calculated from the accumulated
 Faraday cup charge $Q_{F}$ and electron charge $e$. In this experiment $Q_{F}=6$ $\mu$C.  The number of target protons per cm$^2$ is
\begin{equation}
N_{p} = \frac{L_{t} \rho N_{A}}{M_{h}},
\end{equation}
where $L_{t} = 2$~cm is the target length,
$\rho = 0.0708$~g/cm$^{3}$ is the liquid hydrogen density at $T =~20$~K, $N_{A} = 6.02 \times 10^{23}$ is Avogadro's number, and $M_{H} = 1.00794$~g/mol is the atomic mass unit for a natural isotopic mixture of hydrogen. The product $N_{e}N_{p}$ represents the luminosity integrated over time. $A$, $B$, $R$, and $E_{TOF}$ are corrections for acceptance, bin centering, radiative effects and SC efficiency, respectively. $\Delta W$, $\Delta Q^{2}$, $\Delta$cos~$\theta_{\pi^{o}}$, and $\Delta \phi_{\pi^{o}}$ are the bin sizes for the corresponding variables (see Table~\ref{WQ2BinningTable} and Table~\ref{cosThetaPhiBinningTable}). The evaluation of all the factors in the Eq.~(\ref{nEvToCS}) was detailed in the previous sections.

\begin{figure*}
\begin{tabular}{ll}
\includegraphics[scale=0.55]{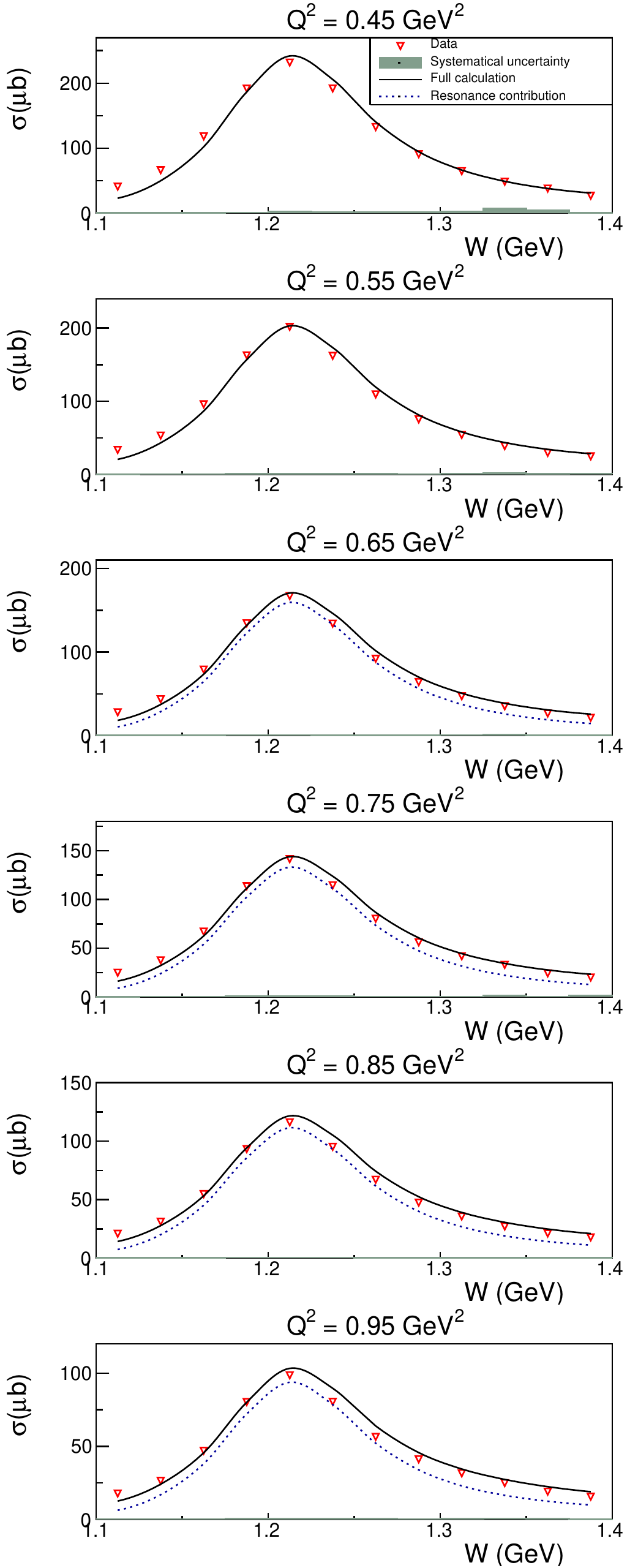}
&
\includegraphics[scale=0.55]{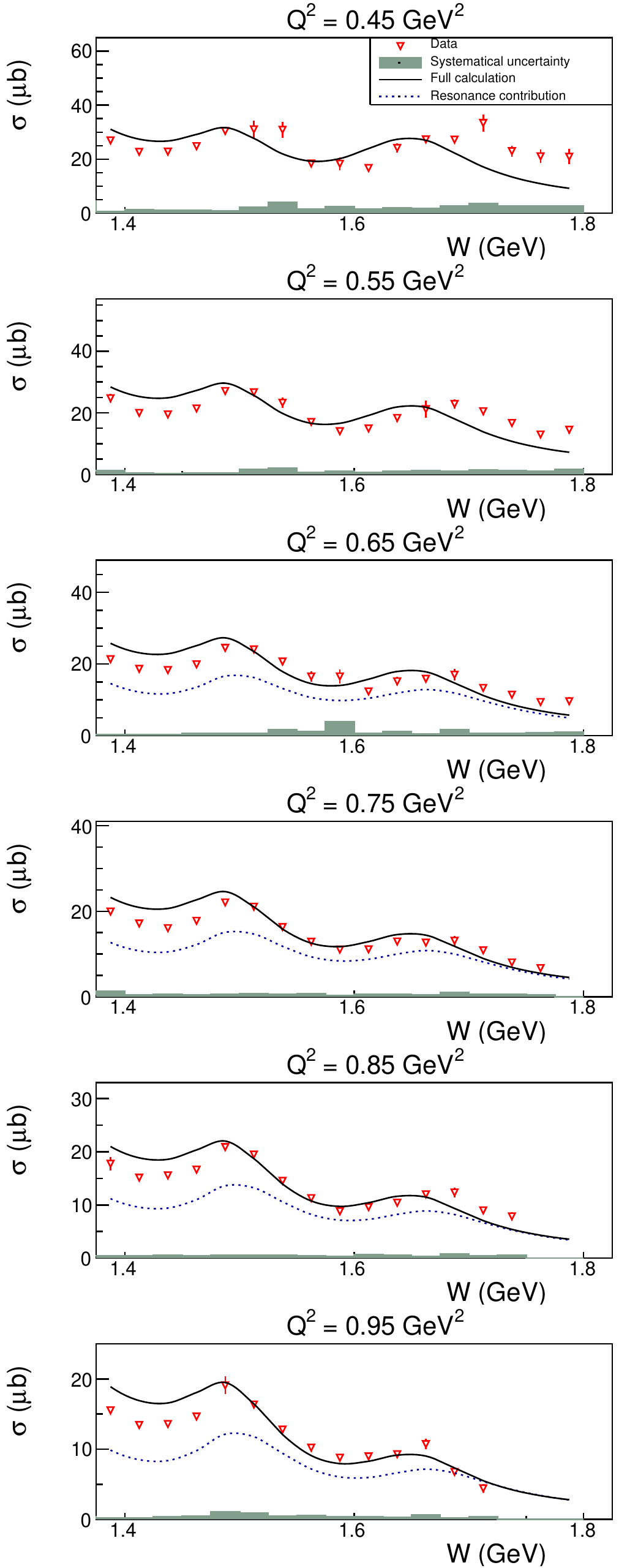}
\end{tabular}
\caption{(Color online) Integrated $\gamma_{v} p \rightarrow \pi^{0} p'$ cross sections as a function of $W$ in the first (left) and second and third (right) resonance regions for different values of $Q^{2}$. The error bars, comparable with the symbol sizes, account for the statistical uncertainties only. Systematic uncertainties are shown by the shadowed areas. Model calculations from the JLab/YerPhi model \cite{JANR} computed using electrocouplings and hadronic decay widths from fits to previous CLAS data \cite{aznauryan, parkAznauryan, mokeev2Pion} are shown as the black solid lines.  The resonance only contributions are shown as the blue dotted lines. The systematic uncertainties are shown by the shadowed areas at the bottom of the plots.}
\label{fullyIntegrated}
\end{figure*}

\begin{figure*}
  \begin{center}
    \epsfig{file =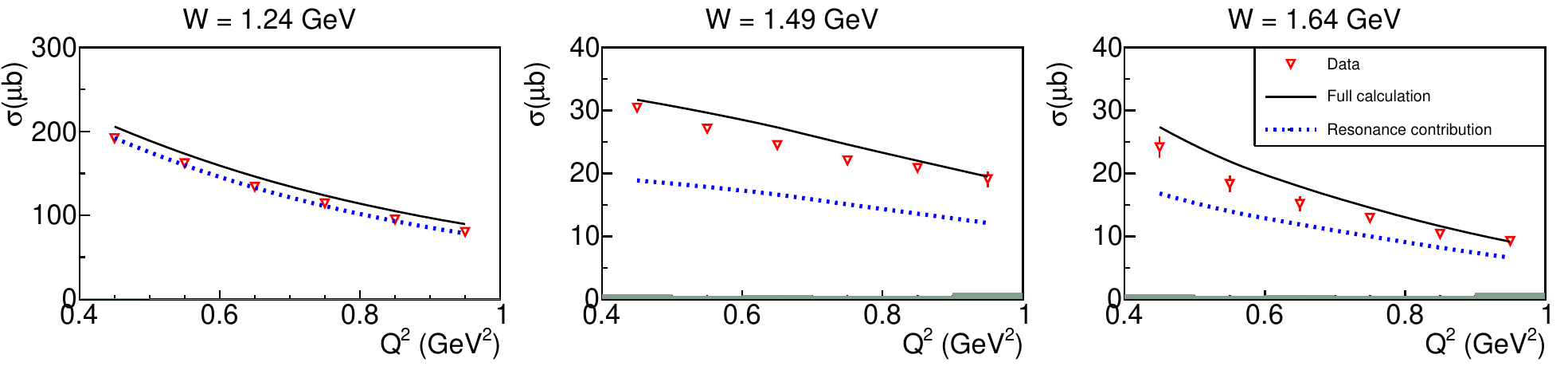, width = 15cm, angle = 0}
    \caption{(Color online) Integrated $\pi^{0}p$ electroproduction cross sections as a function of $Q^{2}$ for selected $W$ bins in the first (left), second (center), and third (right) resonance regions.  Model calculations (full black and resonance only blue dotted lines) are from the JLab/YerPhi model \cite{JANR}. The systematic uncertainties are shown by the shadowed areas at the bottom of the plots.}
    \label{integratedResonance}
    \end{center}
\end{figure*}

The $\gamma_{v} p \rightarrow \pi^{0} p'$ cross sections fully integrated over the center-of-mass angles are shown in Fig.~\ref{fullyIntegrated} as a function of $W$ for all $Q^{2}$ bins used in this measurement.  The $W$ dependence clearly shows three peaks in all $Q^{2}$ bins presented, corresponding to the first, second, and third resonance regions. The model curves shown are predictions based on fits to previous CLAS data.  The first resonance region is dominated by a single isolated state, the $\Delta(1232)3/2^+$, which has been extensively studied over a wide $Q^2$ range. The bump at $W$ $\approx$ 1.5~GeV is dominated by contributions from the $N(1520)3/2^-$ and $N(1535)1/2^-$ states, with much smaller contributions from the Roper $N(1440)1/2^+$ state. Electrocouplings for all of these states were determined by independent studies of the meson electroproduction channels $N\pi$ ~\cite{aznauryan} and $\pi^+\pi^- p$ ~\cite{mokeev2Pion} using proton targets. Similar results for the resonance electrocouplings were obtained from these two channels which have entirely different non-resonant contributions.  This result adds credibility to the self-consistency and model-independence of the analysis ~\cite{mokeevNSTAR}. Currently, the results on the electrocouplings of all resonances with masses less than 1.6~GeV are available in the $Q^2$ range covered so far by our measurements \cite{webIsupov}. 

The $N(1680)5/2^+$ resonance is the most significant contributor to the peak at $W$ $\approx$ 1.7~GeV in the third resonance region. New results on electrocouplings of the $N(1675)5/2^-$, $N(1680)5/2^+$,  and $N(1710)1/2^-$ states have recently become available from analyses of the CLAS $\pi^+n$ electroproduction data in the $Q^2$ range 2.0~GeV$^2$~$<$~$Q^2$~$<~$ 5.0~GeV$^2$~\cite{parkAznauryan}. Our new data will make it possible to determine electrocouplings of the resonances in the third resonance region from the $\pi^0p$ electroproduction channel for the first time at 0.4~GeV~$^2$~$<$~$Q^2$~$<$ 1.0~GeV$^2$.

Finally, the $Q^2$ dependence of $\gamma_{v} p \rightarrow \pi^0p'$ is shown in Fig.~\ref{integratedResonance} for selected $W$ bins in the first, second, and third resonance regions.  The cross sections are well reproduced by the JLab/YerPhi model in the first resonance region, with the $\Delta(1232)3/2^+$ resonance parameters taken from the previous studies. This supports the reliability of our new $\pi^0p$ electroproduction data reported in this paper.  The predicted resonant contributions to the $\pi^0p$ cross section in the second and third resonance regions ranges from significant to dominant. Furthermore, the relative resonance contributions appear to grow with $Q^2$. This feature was also observed in the previous studies of $N\pi$ electroproduction \cite{aznauryan,parkAznauryan}.

\subsection{Exclusive structure functions from $\gamma_{v}p \rightarrow \pi^0p'$ cross sections}\label{strfunct}
The extraction of nucleon resonance electrocouplings for $Q^2>0$~GeV$^{2}$ makes use of both the transverse ($T$) and longitudinal ($L$) polarization states of the virtual photon. These are expressed via the experimental exclusive structure functions $\sigma_T+\epsilon \sigma_L$, $\sigma_{LT}$, and $\sigma_{TT}$,  which can be accessed via the $\phi_{\pi^{0}}$ dependence of the differential $\pi^0p$ cross sections.  Each structure function depends implicitly on $(W, Q^2, \theta_{\pi^{0}})$ and is described by different products of reaction amplitudes and their complex conjugated values \cite{Am72}. The extracted structure functions can also be used to constrain reaction dynamics and non-resonant processes when using model fits to extract resonance parameters. 

To extract the exclusive structure functions from the data, the measured $\dsig$ differential cross sections (see Eq.~\ref{crossSectionFormula}) were fitted in all bins of $(W,Q^2,\theta_{\pi^{0}}, \phi_{\pi^{0}})$ using:
\begin{widetext}
\begin{equation}  
\frac{d\sigma}{d\Omega_{\pi^0}}(W,Q^2,\theta_{\pi^{0}}, \phi_{\pi^{0}}) = A + B\mathrm{cos}{\phi_{\pi^{0}}} + C\mathrm{cos}2\phi_{\pi^{0}}.
\label{strfun_fit}
\end{equation}
\end{widetext}

The fitted coefficients $A, B$, and $C$ are then related to the exclusive structure functions by
\begin{align}
\label{sfFormulas1}
A &= (\sigma_T+\epsilon \sigma_L)\frac{p_{\pi^0}}{k^*_{\gamma}},\\
\label{sfFormulas2}
B &= \sigma_{LT}\frac{p_{\pi^0}}{k^*_{\gamma}} \mathrm{sin}\theta_{\pi^{0}}\sqrt{2\epsilon(\epsilon+1)},\\
\label{sfFormulas3}
C&= \sigma_{TT}\frac{p_{\pi^0}}{k^*_{\gamma}} \mathrm{sin}^2\theta_{\pi^{0}} \epsilon.
\end{align}

 Typical examples of fits to the $\phi_{\pi^{0}}$ dependence of $\dsig$ are shown in Fig.~\ref{secondRegion} along with the resonance contribution to the total cross section.  Examples of the extracted structure functions are shown in Fig.~\ref{strFunctions} and compared to predictions calculated using the resonance electrocouplings and hadronic decay parameters from previous analyses of CLAS data \cite{aznauryan,parkAznauryan,mokeev,mokeev2Pion,mokeev21Pion}. Also shown are the resonant contributions calculated from the JLab/YerPhi model~\cite{JANR}.  Tabulations of all extracted structure functions are available in \cite{clasBD}.

\begin{figure*}[ht]
  \begin{center}
    \epsfig{file =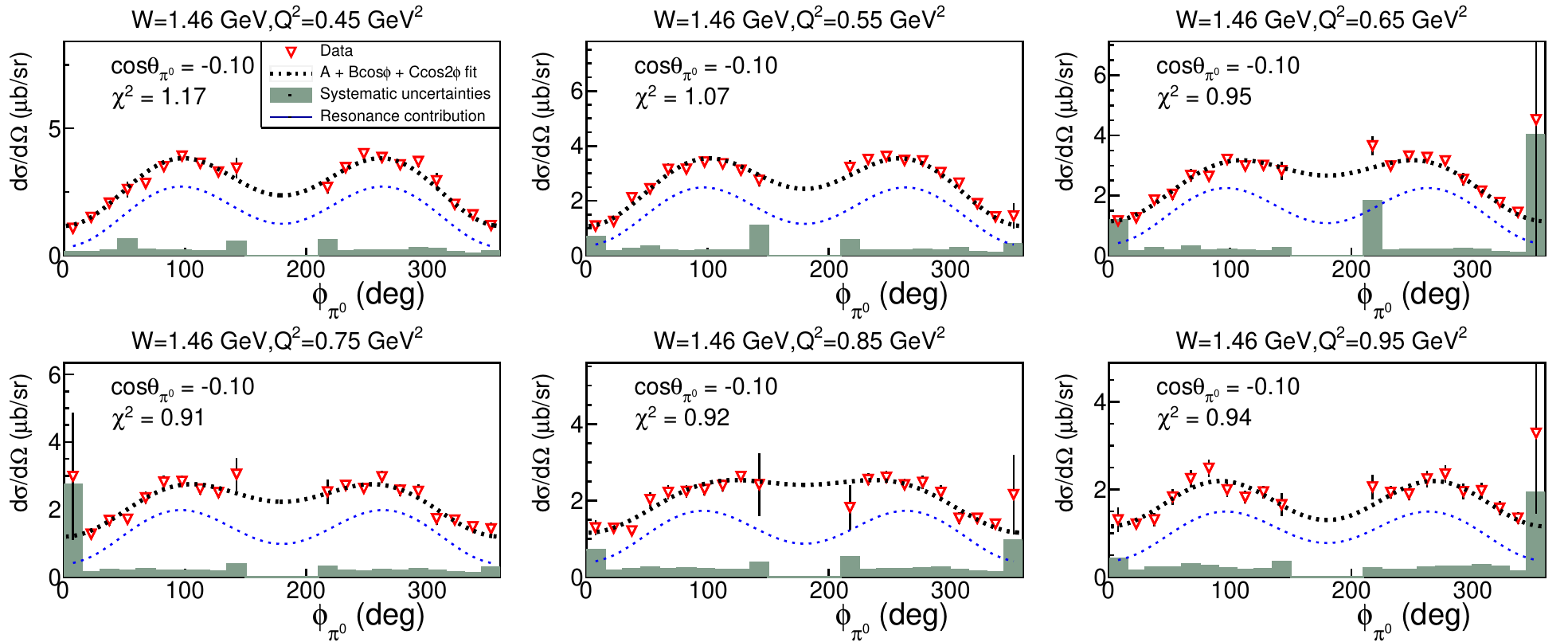, width = 15.5cm, angle = 0}
    \epsfig{file =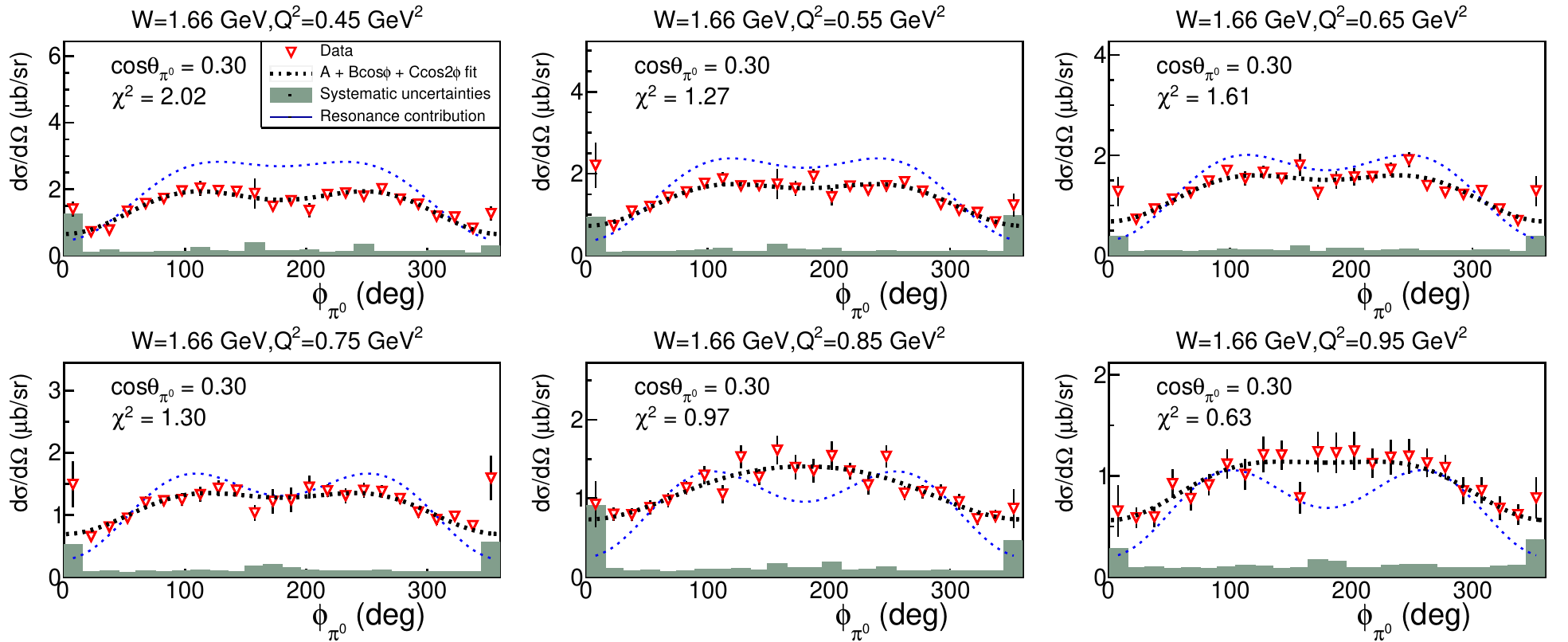, width = 15.5cm, angle = 0}
    \caption{(Color online) Cross sections $\dsig$ as a function of the center-of-mass angle $\phi_{\pi^{0}}$ in different bins of (W, $Q^{2}$, cos$\theta_{\pi^{0}}$). The fits using Eq. (\ref{strfun_fit}) are shown by the thick black dashed lines. The fit $\chi^{2}$ are listed in the respective panels. The dashed blue lines represent the resonance contributions calculated from the JLab/YerPhi model~\cite{JANR}. Shaded bands represent systematic uncertainty.}
    \label{secondRegion}
    \end{center}
\end{figure*}

\begin{figure*}[ht]
  \begin{center}
    \epsfig{file =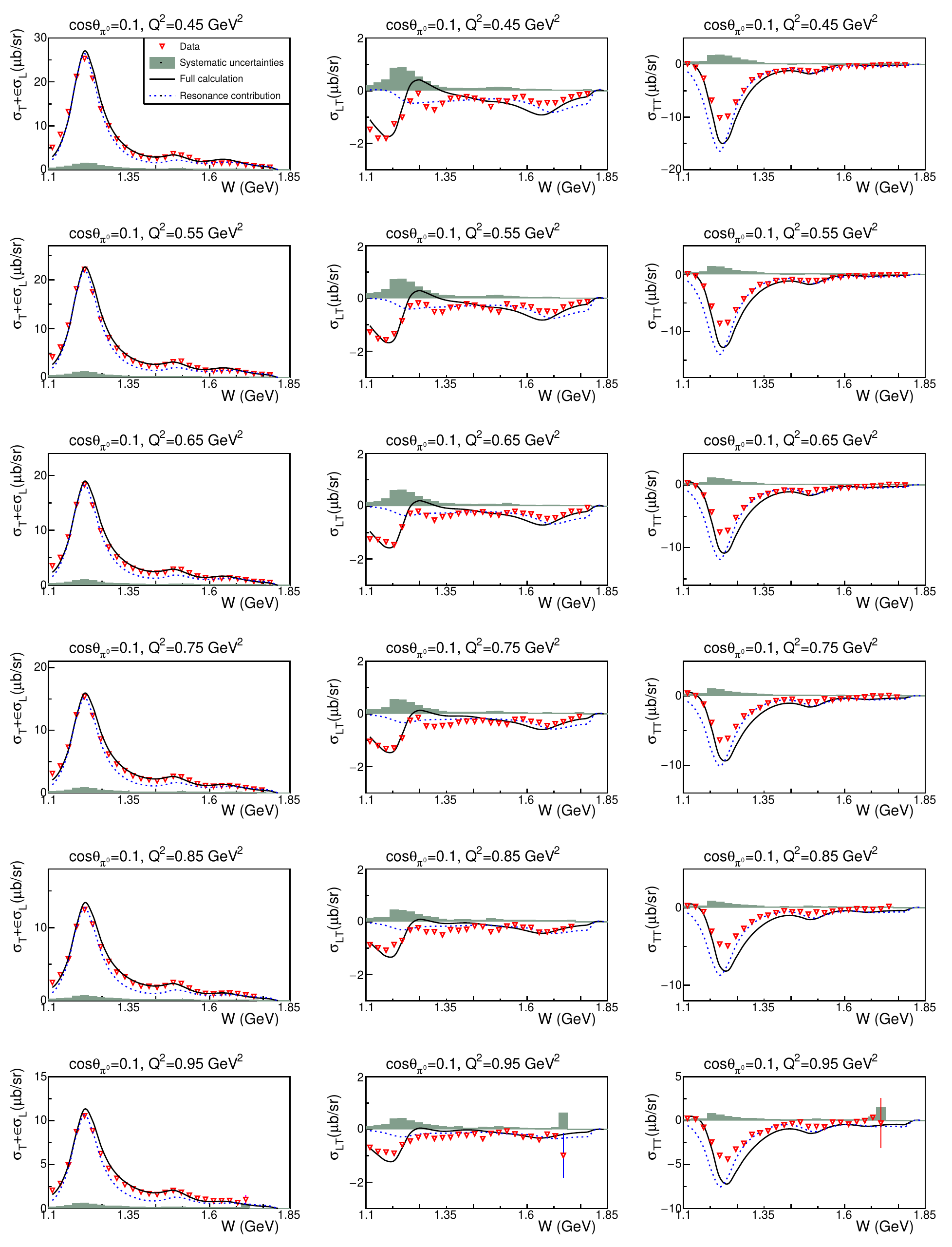, width = 14.5cm, angle = 0}
    \caption{(Color online) $W$ dependencies of the exclusive structure functions $\sigma_T+\epsilon\sigma_{L}$,  $\sigma_{LT}$, and  $\sigma_{TT}$ in different bins of the (cos$\theta_{\pi^{0}}$, $Q^{2}$). Computation of the exclusive structure functions is done within the framework of the JLab/YerPhi model~\cite{JANR} and with the resonance parameters determined from the CLAS exclusive meson electroproduction data \cite{aznauryan,parkAznauryan,mokeev,mokeev2Pion,mokeev21Pion} and are shown by the solid lines, while the blue dashed lines represent the resonant contributions.  Shaded bands represent systematic uncertainty.}
    \label{strFunctions}
    \end{center}
\end{figure*}
\subsection{Legendre multipole expansion of the structure functions}
A Legendre multipole expansion of the structure functions can reveal the partial wave composition of the $\gamma_{v}p \rightarrow \pi^0p$ reaction. $N\pi$ decays of the resonances of a particular spin-parity produce in the final state well defined set of the pion orbital angular momentum $l_{\pi}$. Since the partial wave for the $\gamma_{v}p \rightarrow \pi^0p$ reaction also corresponds to the certain set of $l_{\pi}$, analysis of the Legendre moments can enhance the possible signatures of nucleon resonances in the experimental data.

The general form of the expansion can be expressed by
\begin{align}
 \sigma_{T} + \epsilon\sigma_{L} &= \sum_{i = 0}^{2l}A_{i}P_{i}(cos\theta^{*}_{\pi})\label{legendreAGeneral},\\
\sigma_{LT} &= \sum_{i = 0}^{2l -1}C_{i}P_{i}(cos\theta_{\pi}^{*})\label{legendreCGeneral}, and\\
\sigma_{TT} &= \sum_{i = 0}^{2l - 2}B_{i}P_{i}(cos\theta^{*}_{\pi})\label{legendreBGeneral},
\end{align}
where $l$ is the maximal orbital momentum of the $\pi^{0}p$ final states in the truncated expansion. Each coefficient in Eqs.~(\ref{legendreAGeneral}-\ref{legendreBGeneral}) can be in turn related to electromagnetic multipoles El, Ml, and Ll~\cite{raskinDonnelly, aznaryanBurkert}. In order to obtain from our data the input for the partial wave analyses, we performed a decomposition of the structure functions for $\pi^0 p$ electroproduction over sets of Legendre multipoles. We restricted the $\pi^0p$ relative orbital momentum l~$\leq$~3. Representative examples of the Legendre multipoles are shown in Fig.~\ref{legendreAllRegionCoeff}. Numerical results on Legendre multipoles determined from our data are available in the CLAS Physics Data Base~\cite{clasBD}. The $W$-dependencies of $A_{0}$ and $B_{2}$ Legendre multipoles demonstrate resonance-like structure at $W$ around 1.68~GeV in the entire $Q^2$ range covered in our measurements. In the $W$-interval from 1.5~GeV to 1.65~GeV, the Legendre multipoles $C_{1}$ and $A_{2}$ decreases and increases with $W$, respectively, while at $W$ $>$ 1.65~GeV they become almost $W$-independent. These features were observed in all $Q^2$-bins covered by our data.

\subsection{Resonance contributions}

For preliminary studies of the resonance contributions from the experimental data of our paper, we computed the integrated and differential $\pi^0p$ cross sections, exclusive structure functions and their Legendre moments within the JLab/YerPhI amplitude analysis framework \cite{JANR}. It incorporates two different approaches: unitary isobar model and fixed-t dispersion relation allowing us to compute full $\gamma_{v}p \rightarrow N\pi$ electroproduction off proton amplitudes by fitting to data the nucleon resonance parameters only, while the parameters of the non-resonant contributions are taken from analyses of other experiments and fixed within their uncertainties. The JLab/YerPhI amplitude analysis framework provided the dominant part of the  worldwide available information on resonance electrocouplings from exclusive $N\pi$ electroproduction off protons \cite{aznauryan,parkAznauryan,aznaryanBurkert}. In the computations of the observables presented here, we used nucleon resonance electrocouplings available from the analyses of the CLAS results on exclusive $N\pi$, $p\eta$, and $\pi^{+}\pi^{-}p$ electroproduction off protons~\cite{mokeevNSTAR} and stored in the web \cite{webIsupov}.  The resonance hadronic decay parameters were taken from \cite{aznauryan,parkAznauryan,mokeev}. A list of the resonances included in the description of the $\pi^0p$ data is shown in Table~\ref{resonances} together with 
their total widths and branching fractions for decays to the $\pi^0p$ final state.
\begin{table}
\centering
\begin{tabular}{|c|c|c|c|c|c|}
\hline Resonance&Width, MeV&Branching ratio to  \\& & $\pi^{0}$p channel, \%\\
\hline $\Delta(1232)\frac{3}{2}^{+}$& 115& 65 \%\\
\hline $N(1535)\frac{1}{2}^{-}$& 150& 15 \%\\
\hline $N(1440)\frac{1}{2}^{+}$& 350& 20 \%\\
\hline $N(1520)\frac{3}{2}^{-}$&115&20 \%\\
\hline $N(1650)\frac{1}{2}^{-}$&140&25 \%\\
\hline $N(1675)\frac{5}{2}^{-}$&150&15 \%\\
\hline $N(1680)\frac{5}{2}^{+}$&130&20 \%\\
\hline $\Delta(1600)\frac{3}{2}^{+}$&320&15 \%\\
\hline $\Delta(1620)\frac{1}{2}^{-}$&140&20 \%\\
\hline $\Delta(1700)\frac{3}{2}^{-}$&300&15 \%\\
\hline
\end{tabular}
\caption{The nucleon resonances included into the JLab/YerPhI approach~\cite{JANR} in the description of exclusive $ep \to e'p'\pi^{0}$ electroproduction channel.}
\label{resonances}
\end{table}

The evaluations of exclusive structure functions within the JLAB/YerPhi~\cite{JANR} amplitude analysis framework with resonance parameters from the exclusive CLAS electroproduction data \cite{aznauryan,parkAznauryan,mokeev,mokeev2Pion,mokeev21Pion} are shown in Fig.~\ref{strFunctions} by solid lines, while the resonant contributions are shown by dashed lines. The close description of our data on fully integrated and differential cross sections(Figs.~\ref{fullyIntegrated},~\ref{integratedResonance},~\ref{secondRegion}), exclusive structure functions (Fig.~\ref{strFunctions}) was achieved without adjustment of the resonant and non-resonant parameters and demonstrated the large resonant contributions into $\pi^0p$ electroproduction off protons in the second and the third resonance regions. We further investigated the data sensitivity to the variation of the electrocouplings  of excited nucleon states in the third resonance region. 

\begin{figure*}[h]
  \begin{center}
    \epsfig{file =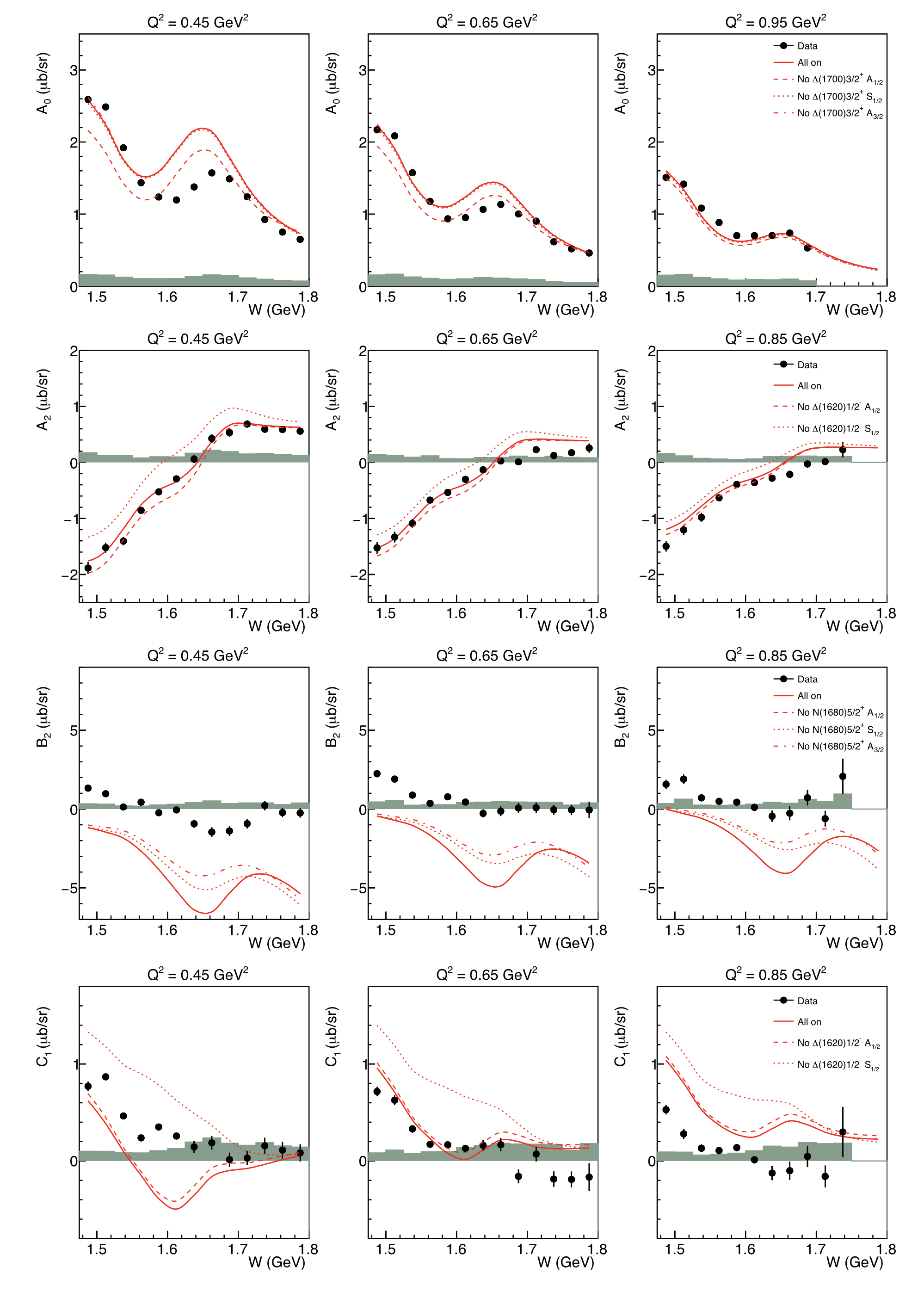, width = 14cm, angle = 0}
    \caption{(Color online) Representative Legendre moments at different photon virtualities $Q^2$ as the functions of $W$ in comparison with the JANR/YerPhi model expectations~\cite{JANR} with the electrocouplings of the different resonances turned on/off. From top to bottom: $A_{0}$ and manifestation of the sensitivity to the $\Delta$(1700)3/2$^+$, $A_{2}$ and manifestation of the sensitivity to the $\Delta$(1620)1/2$^-$, $B_{2}$ and manifestation of the sensitivity to the $N$(1680)5/2$^+$, $C_{1}$ and manifestation of the sensitivity to the $\Delta$(1620)1/2$^-$. Shaded bands represent systematic uncertainty.}
    \label{legendreAllRegionCoeff}
    \end{center}
\end{figure*}
\begin{figure*}[h]
  \begin{center}
    \epsfig{file =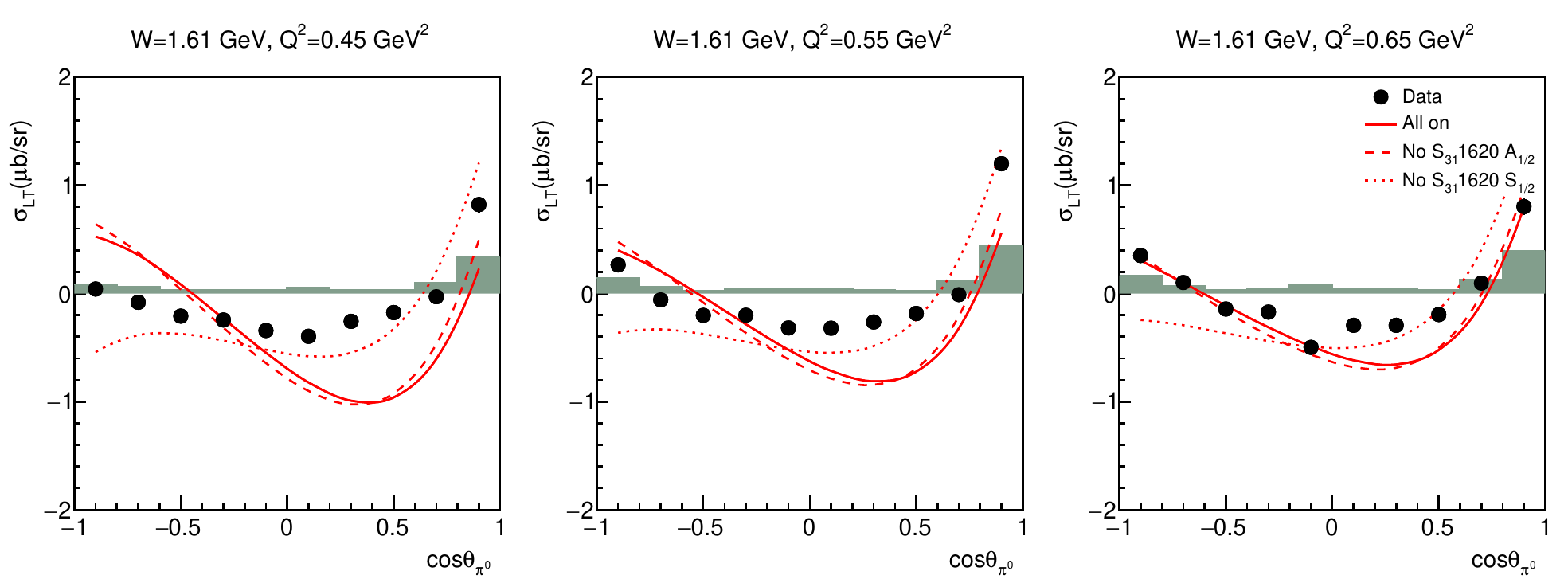, width = 15cm, angle = 0}
    \caption{(Color online) The $\sigma_{LT}$ structure function at $W=1.61$~GeV and different photon virtualities $Q^2$  as the functions of $cos(\theta^{*}_{\pi^0})$ CM angles in comparison with the JLAB/YerPhi approach expectations~\cite{JANR}  with turned on/off electrocouplings of the $\Delta(1620)1/2^-$ resonance: all electrocouplings on (solid lines) $A_{1/2}$ electrocoupling off (dashed lines), and $S_{1/2}$ electrocoupling off (dotted lines). Shaded bands represent systematic uncertainty.}
    \label{legendreThirdRegionSFC}
    \end{center}
\end{figure*}

\subsection{Manifestations of individual resonances in the $\pi^{0}p$ electroproduction observables}
So far, the most detailed information on the $Q^2$ evolution of the resonance electrocouplings is available for the $\Delta(1232)3/2^+$ resonance and for the excited nucleon states in the second resonance region. Our data will extend the results on nucleon resonance electrocouplings into the third resonance region. 

Resonances with $I=3/2$ couple preferentially to the $\pi^0p$ final state, due to isospin conservation.  Although the $I=3/2$ states $\Delta(1620)1/2^-$ and $\Delta(1700)3/2^-$ are located in third resonance region, their contributions to the fully integrated cross sections are rather small. The resonant part is clearly dominated by the contributions from the $I=1/2$ states $N(1520)3/2^-$, $N(1535)1/2^-$, and $N(1680)5/2^+$. It is known that the $\Delta(1620)1/2^-$ and $\Delta(1700)3/2^-$ resonances decay preferentially via $N\pi\pi$, and in particular the $\pi^+\pi^-p$ channel is the primary source of information on these electrocouplings. The results on electrocouplings of the $\Delta(1620)1/2^-$ and $\Delta(1700)3/2^-$ resonances from $\pi^+\pi^-p$  photoproduction \cite{Gol18} and electroproduction \cite{mokeev21Pion,mokeev} have already become available. 

Improving our knowledge of these $I=3/2$ states from studies of $\pi^0p$ electroproduction, with completely different non-resonant contributions in comparison to the $\pi^+\pi^-p$ exclusive channel, is of particular importance in order to further test the model dependence of the extraction of the fundamental resonance electrocouplings. As a preliminary exercise we checked the sensitivity of our measured observables to contributions from the $\Delta(1620)1/2^-$ and $\Delta(1700)3/2^-$ resonances by turning on/off particular electrocouplings of these states using the JLab/YerPhI amplitude analysis framework. Observed discrepancy between data and  computations in the third resonance region is due to the lack of the previously available data. We will need a comprehensive analysis of the newly available data for sound evaluation of both the resonance and background contribution to the cross section.

\begin{figure*}
  \begin{center}
    \epsfig{file =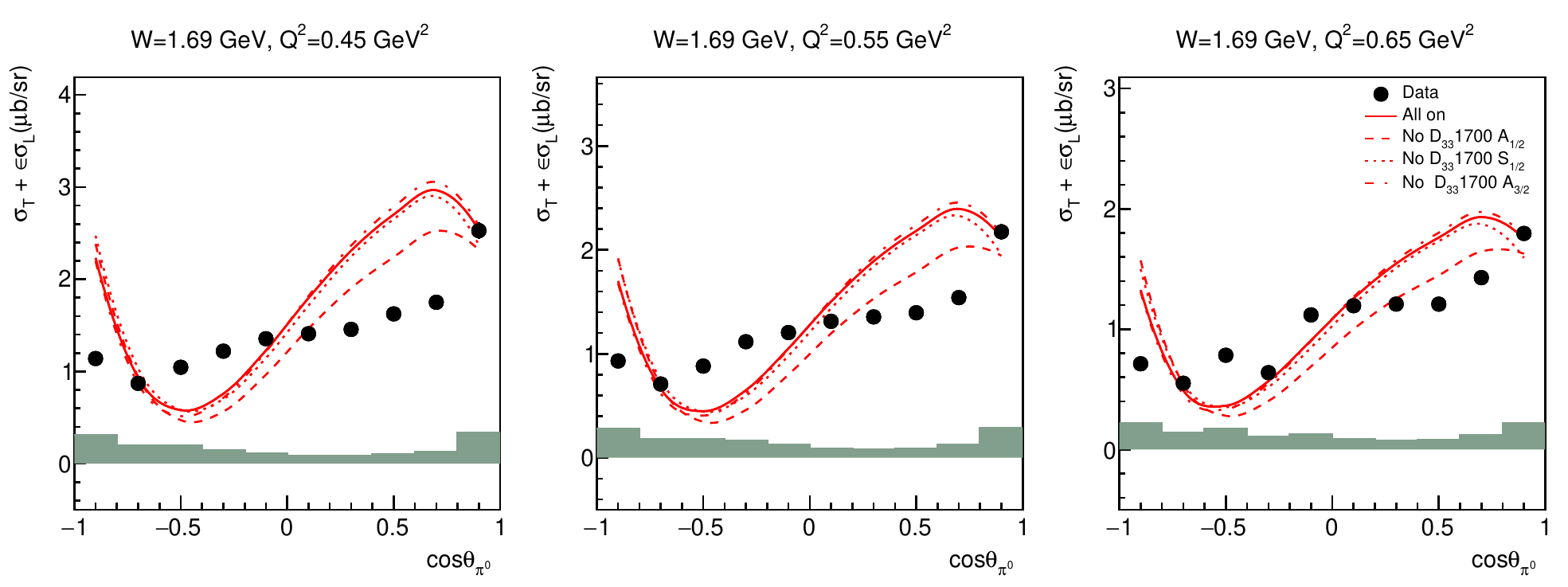, width = 15cm, angle = 0}
    \caption{(Color online) $\sigma_{T} + \epsilon\sigma_{L}$ unpolarized structure function at $W=1.69$~GeV and different photon virtualities $Q^2$  as the functions of $cos(\theta^{*}_{\pi^0})$ CM angles in comparison with the JLab/YerPhi model expectations~\cite{JANR}  with turned on/off electrocouplings of $\Delta(1700)3/2^-$ resonance: all electrocouplings on (solid lines), $A_{1/2}$ electrocoupling off (dashed lines), $S_{1/2}$ electrocoupling off (dotted lines), $A_{3/2}$ electrocoupling off (dash-dotted lines). Shaded bands represent systematic uncertainty.}
    \label{legendreThirdRegionSFA}
    \end{center}
\end{figure*}

\begin{figure*}
  \begin{center}
    \epsfig{file =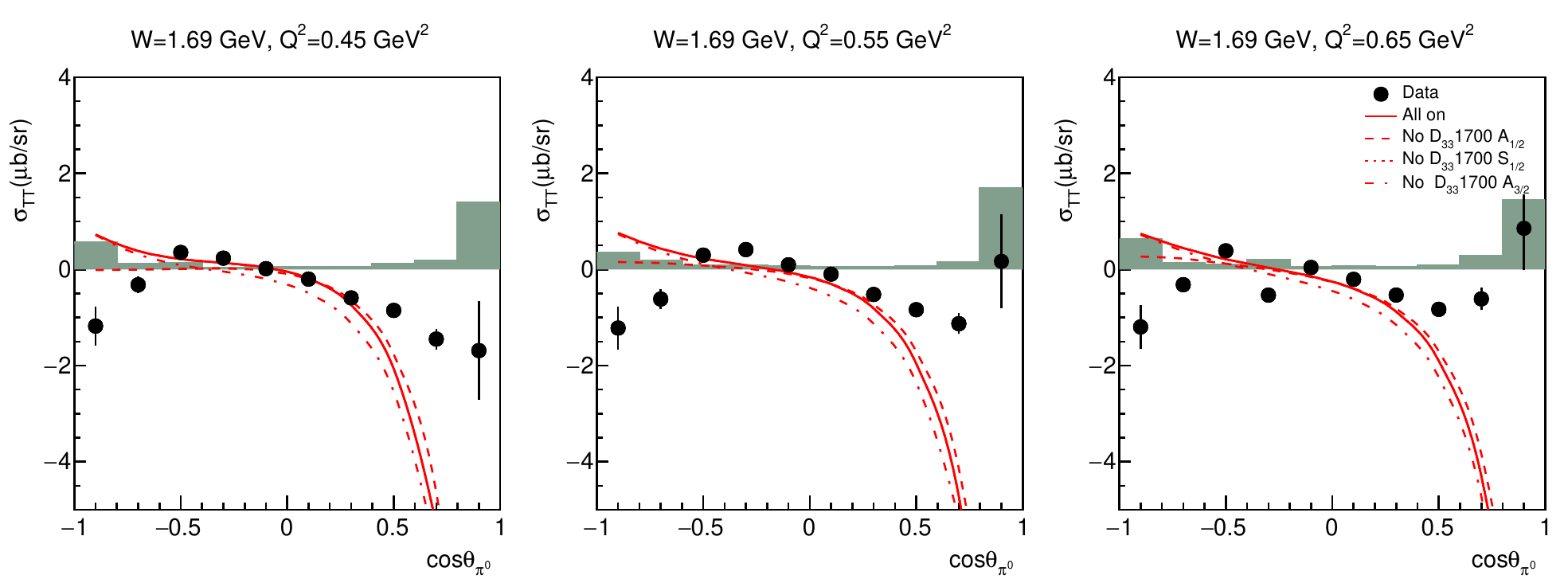, width = 15cm, angle = 0}
    \caption{(Color online) $\sigma_{TT}$ unpolarized structure function at $W=1.69$~GeV and different photon virtualities $Q^2$  as the functions of $cos(\theta^{*}_{\pi^0})$ CM angles in comparison with the JLab/YerPhi model expectations~\cite{JANR}  with turned on/off electrocouplings of $\Delta(1700)3/2^-$ resonance: all electrocouplings on (solid lines), $A_{1/2}$ electrocoupling off (dashed lines), $S_{1/2}$ electrocoupling off (dotted lines), $A_{3/2}$ electrocoupling off (dash-dotted lines). Shaded bands represent systematic uncertainty.}
    \label{legendreThirdRegionSFB}
    \end{center}
\end{figure*}

The $\Delta(1620)1/2^-$ resonance is the only known state with a dominant longitudinal $S_{1/2}$ coupling in the $Q^2$ range 0.5-1.5~GeV$^2$. Sensitivity to this state can be demonstrated in the angular dependence of the longitudinal-transverse $\sigma_{LT}$ structure function (Fig.~\ref{legendreThirdRegionSFC}) at $W$ near the resonant point and in the $W$ dependence of the $C_1$ Legendre moment (Fig.~\ref{legendreAllRegionCoeff}). Both observables show significant sensitivity to the $S_{1/2}$ electrocoupling, where the difference between the computed observables with $S_{1/2}$ electrocoupling turned on/off is far outside of the range of systematical uncertainties for the data. Electrocouplings for this state obtained from the analysis \cite{mokeev21Pion} of the CLAS $\pi^+\pi^-p$ electroproduction data \cite{Ri03} showed the biggest contributions from longitudinal amplitudes to the electroexcitation of this state at 0.5~GeV$^2$ $<$~$Q^2$~$<$~1.5~GeV$^2$.\\

The $\Delta(1700)3/2^-$ state is not visible in the $W$ dependence of $\dsig$ shown in Fig.~\ref{fullyIntegrated} because of the large value of the total decay width (Table~\ref{resonances}). Therefore, the extraction of the $\Delta(1700)3/2^-$ electrocouplings requires a partial wave analysis of the extracted structure functions. Both the angular dependence of $\sigma_{T} + \epsilon\sigma_{L}$(Fig.~\ref{legendreThirdRegionSFA}) and the $A_{0}$ Legendre moment (Fig.~\ref{legendreAllRegionCoeff}) demonstrate the sensitivity of these observables to the $A_{1/2}$ electroexcitation amplitudes of the $\Delta(1700)3/2^-$ resonance. On the other hand, the angular dependence of $\sigma_{TT}$ near the resonant point are sensitive to the $A_{3/2}$ electrocouplings as shown in Fig.~\ref{legendreThirdRegionSFB}. Moreover, the significant differences in the behavior of the computed $\sigma_{TT}$ structure functions and our data at small pion CM emission angles suggest the need for the further studies of resonant and non-resonant amplitudes in this kinematic region.\\

According to the results in Fig.~\ref{legendreAllRegionCoeff}, Legendre moment $B_{2}$ demonstrates strong sensitivity to the contribution from $N(1680)5/2^+$ state. Therefore, the combined studies of $\pi^0p$ and $\pi^+n$ electroproduction off protons are of particular importance for extension of the results on this state electrocouplings and verification of their consistency from analyses of different single-pion electroproduction off proton channels.

 \section{Summary}
High statistics measurements of the $ep \to e'p' \pi^{0}$ exclusive channel in the W range from 1.1 to 1.8~GeV and photon virtualities $Q^{2}$  from 0.4 to 1.0~GeV$^{2}$ with nearly complete angular coverage are presented. For the first time, experimental data on this exclusive channel in the aforementioned kinematics have become available. Two-fold differential $\dsig$ and fully integrated cross sections are measured with unprecedented accuracy. Unpolarized structure functions $\sigma_T+\epsilon\sigma_L$ and the interference longitudinal-transverse $\sigma_{LT}$ and transverse-transverse $\sigma_{TT}$ structure functions are extracted from fits to the $\phi^*_{\pi^{0}}$ dependence, and their Legendre moments are evaluated.

Phenomenological analysis of these results within the JLab/YerPhI amplitude analysis framework \cite{JANR}, using resonance parameters from fits to previous exclusive CLAS electroproduction data \cite{aznauryan,parkAznauryan,mokeev,mokeev2Pion,mokeev21Pion}, reveal sensitivity to resonant contributions in the entire kinematic area covered by our measurements. Furthermore, an approximate description of the new $\pi^0p$ data with the JLAB/YerPhI model is seen using these resonance parameters. These observations are a good indication of the possibility of the extraction of the electroexcitation amplitudes of the nucleon resonances in the third resonance mass range $W$ $>$ 1.6~GeV in the $\pi^{0}p$ channel at 0.4~$\le$~$Q^2$~$\le$~1.0~GeV$^2$. They can be compared with the already available electrocouplings for the excited states in the third resonance region as determined from the CLAS $\pi^+\pi^-p$ electroproduction data \cite{mokeev,mokeev21Pion}.

Isospin Clebsch-Gordan coefficients imply preferential decays of isospin 3/2  $\Delta$ resonances to the $\pi^0p$ final state. In fact the two lightest of the $\Delta^*$ states in the third resonance region, $\Delta(1620)1/2^-$ and $\Delta(1700)3/2^-$, decay preferentially to the  $N\pi\pi$ final states, with the $\pi^+\pi^-p$ electroproduction channel providing the major source of the information on theses states. However the exclusive $\pi^0p$ structure functions and their Legendre moments demonstrate also sizable sensitivity to the electrocouplings of the $\Delta(1620)1/2^-$ and $\Delta(1700)3/2^-$ resonances. The results on these electrocouplings from $\pi^0p$ channel will be essential in order to support their extraction from the $\pi^+\pi^-p$ electroproduction observables in a nearly model-independent way. A new opportunity to verify consistency of resonance electrocoupling extraction from independent studies of $\pi^0p$ and $\pi^+\pi^-p$ electroproduction channels was recently provided by the new CLAS data on $\pi^+\pi^-p$ electroproduction cross sections \cite{Fe18} obtained in the same range of $W$ and $Q^2$ and from the same experimental run as the $\pi^0p$ data presented in this paper. The results on electrocoupling of the high-lying excited nucleon states will improve the knowledge on the resonant contributions into inclusive electron scattering observables estimated within the  approach~\cite{Ad1}. Credible evaluation of the resonant contributions into inclusive electron scattering opens up new opportunities for the insight into the ground nucleon parton distributions at large $x$-Bjorken and for exploration of quark-hadron duality.

 \section{Acknowledgements}
We thank the staff of the Accelerator and Physics Divisions at Jefferson Lab for making the experiment possible. We are grateful to I. G. Aznauryan, C. D. Roberts, E. Santopinto, and J. Segovia for helpful discussions. This work was supported in part by the U.S. Department of Energy (DE-FG-04ER41309), the National Science Foundation, the Skobeltsyn Institute of Nuclear Physics and the Physics Department at Moscow State University, the French Centre National de la Recherche Scientifique and Commissariat ˆ lÕEnergie Atomique, the French-American Cultural Exchange (FACE), the Italian Instituto Nazionale di Fisica Nucleare (INFN), the Chilean Comisi—n Nacional de Investigaci—n Cient'fica y Tecnol—gica (CON-ICYT), the National Research Foundation of Korea, and the UK Science and Technology Facilities Council (STFC). Jefferson Science Associates (JSA) operates the Thomas Jefferson National Accelerator Facility under Contract DE-AC05-06OR23177.

\end{document}